\documentclass[aps,amsmath,amssymb,twocolumn,showpacs,prx]{revtex4-1}

\usepackage{graphicx} 
\usepackage{dcolumn}
\usepackage{bm}
\usepackage{amsmath}
\usepackage{amssymb}
\usepackage{color}
\usepackage{bbold}
\usepackage{bm}

\newcommand{\cl}[1]{\hat{\mathcal{#1}}}

\newcommand{\ket}[1]{|#1\rangle}
\newcommand{\bra}[1]{\langle#1|}
\newcommand{\ave}[1]{\langle #1 \rangle}

\newcommand{\tr}{\text{Tr}}

\newcommand{\llangle}[1][]{\savebox{\@brx}{\(\m@th{#1\langle}\)}%
  \mathopen{\copy\@brx\kern-0.5\wd\@brx\usebox{\@brx}}}
\newcommand{\rrangle}[1][]{\savebox{\@brx}{\(\m@th{#1\rangle}\)}%
  \mathclose{\copy\@brx\kern-0.5\wd\@brx\usebox{\@brx}}}

\begin{document}
\title{Time-dependent generalized Gibbs ensembles in open quantum systems}
\author{Florian Lange}
\author{Zala Lenar\v{c}i\v{c}}
\author{Achim Rosch}
\affiliation{Institute for Theoretical Physics, University of Cologne, Z\"ulpicher Stra{\ss}e 77a, D-50937 Cologne, Germany}

\begin{abstract}
Generalized Gibbs ensembles have been used as powerful tools to describe the steady state of  integrable many-particle quantum systems after 
a sudden change of the Hamiltonian. Here we demonstrate numerically, that they can be used for a much broader class of problems. We consider
integrable systems in the presence of weak perturbations  which both break integrability and drive the system to a state far from equilibrium. Under these conditions, we show that the steady state and the time-evolution on long time-scales can be accurately described  by  a (truncated) 
generalized Gibbs ensemble with time-dependent Lagrange parameters, determined from simple rate equations. We compare the  numerically exact
time evolutions of density matrices for small systems with a theory based on block-diagonal density matrices (diagonal ensemble) and a time-dependent generalized Gibbs ensemble containing only  small number of approximately conserved quantities, using
the one-dimensional Heisenberg model with perturbations described by Lindblad operators as an example.
\end{abstract}


\maketitle

\section{Introduction}
In recent years the thermalization of closed quantum systems has been intensively studied \cite{polkovnikov11,gogolin16,dalessio016}. A typical setup is the quantum quench: a system, initialized in the ground state of a Hamiltonian $H_i$, undergoes a non-equilibrium dynamics due to evolution with another Hamiltonian $H_f$. In the long time limit generic ergodic many-particle systems are expected to reach a thermal Gibbs state, $\rho \sim e^{-\beta H_f}$ \cite{rigol08}. The approach to this state can, however, be slow and is generically characterized by power-laws \cite{lux14,bohrdt17,leviatan17}. 

In integrable systems, in contrast, the existence of macroscopically many local conserved quantities restricts the dynamics and prohibits conventional thermalization \cite{essler16,vidmar16,calabrese16,cazalilla16}.
It has, however, been conjectured \cite{rigol07} that generalized Gibbs ensembles (GGEs) provide an accurate thermodynamic description of steady states in this case, at least for local observables. For each (quasi-)local conserved quantity a Lagrange parameter is introduced, generalizing the concept of temperature or chemical potential. The Lagrange parameters are thereby determined from the value of conserved quantities in the initial state. Exact local equivalence of GGEs and diagonal ensembles can be proven only after a complete set of local conserved quantities is included into the GGE, and is for non-interacting models equivalent to including all mode occupation numbers \cite{rigol07,fagotti13gge2}. For the Heisenberg model, as a prototypical interacting quantum integrable model, quenches have been fist studied by \cite{pozsgay13,fagotti13}. A vast improvement in agreement has been achieved \cite{ilievski15a} only after the discovery of additional quasi-local conserved quantities \cite{ilievski15,mierzejewski15,ilievski16}, by systematically including families of quasi-local conservation laws. Using the quenched action approach \cite{caux16,caux13}, it was possible to demonstrate that GGEs indeed destribe the steady state at least for certain initial conditions \cite{brockmann14gge,wouters14,pozsgay14,mestyan15gge,piroli16gge}. 
It has also been suggested that convergent approximations to the steady state are obtained when more and more conservation laws are included \cite{pozsgay17}.

Using ultracold atoms integrable models can be realized with such a high precision, that one can neglect integrability breaking terms 
at least up to some time \cite{kinoshita06,gring12,smith13,Langen2016}. In spectacular experiments, Langen {\it et al.}  \cite{langen15} succeeded to demonstrate that a truncated GGE can describe high-order steady-state correlation functions of an interacting Bose gas after a quench with a high precision. Integrability breaking perturbations can also be controllably tuned \cite{tang17} to display the crossover from integrable dynamics to thermalization. 

In condensed matter systems, one can realize approximately integrable systems for example in spin-chain materials. In this case, however, one cannot neglect integrability breaking terms arising, e.g., from phonons, intra-chain coupling or other terms not described by simple spin-$1/2$ one-dimensional Heisenberg model. Due to the proximity to integrable points, the heat conductivity in such systems can be strongly enhanced \cite{jung06,jung07}, but the steady state is expected to become thermal after a quantum quench. 
Nevertheless, it was shown that a static weak integrability breaking induces thermalization only on the longest time scale \cite{bertini15,eckstein09,stark13,Marino2012,Marcuzzi2013,bertini16,buchhold16,biebl17}, while the transient dynamics dwells on the so-called prethermal plateaux \cite{moeckel08,kollar11}
that can be described by a GGE with Lagrange multipliers fixed by the initial state using approximately conserved quantities, possibly perturbatively readjusted according to the weak integrability breaking \cite{essler14}.

After a quantum quench, all exotic concerved quantities decay in the presence of integrability breaking perturbation.
The situation is, however, completely different for another class of problems: (weakly) driven systems.
These are systems,  where time-dependent perturbations or the coupling
to non-thermal reservoirs drive the system towards a non-equilibrium state. For example, we considered in Ref.~\cite{lange17}
the coupling of a spin-chain to laser light and phonons.
In this case, the decay of conserved quantities due to integrability breaking can be balanced by gain terms arising
from the driving terms. Generically, a macroscopic set of approximate conservation laws is activated despite the presence of integrability breaking. Contrary to the case of quench problems, the value of the conserved quantities in the steady state is not determined by initial conditions but by the balance of driving terms, the coupling to thermal and non-thermal baths, and other integrability breaking terms.
We argued in \cite{lange17} that the resulting states are far from equilibrium and one can use a GGE to describe them quantitatively as long as all driving and integrabilty breaking terms are weak. We showed that one can use this ideas to realize novel types of spin or heat pumps.

The main goal of the present paper is to demonstrate numerically that GGEs with time-dependent Lagrange parameters accurately
describe both the time-evolution and the steady state of weakly driven approximately integrable many-particle quantum systems.
As an example we chose a one-dimensional spin-$1/2$ Heisenberg model coupled to a non-thermal bath described by Lindblad operators.
We have chosen this model because it is better suited for numerical analysis compared to the much more complicated case considered in Ref.~\cite{lange17}. Our approach based on time-dependent GGEs covers both the physics of prethermalization and the relaxation towards a steady state potentially far from thermal equilibrium.
The concept of time-dependent GGE has been suggested earlier, to our knowledge in Refs.~\cite{stark13,fagotti14tdep,bertini15a}. Ref.~\cite{stark13} uses it only implicitly while constructing an effective quantum Boltzmann equation that captures prethermal-to-thermal regime for weakly interacting systems (a Boltzmann equation was also analyzed by us in Ref.~\cite{genske15,lenarcic18}). Ref.~\cite{fagotti14tdep} addresses quenches from  superintegrable (containing additional symmetries) to post-quench noninteracting integrable models (that weakly break those symmetries). Observed prethermalization and subsequent equilibration to a GGE spanned by the conservation laws of the final model was also captured by a time-dependent GGE. Unlike in our setup, in this case integrability is preserved throughout the whole evolution and the time-dependence arises because the additional symmetries do not commute with the local conserved quantities of the final Hamiltonian. A similar condition also applies  for \cite{bertini15a}, which studied weakly interacting integrable models with and without weak integrability breaking and showed that mean-field equation capture the dynamics at intermediate times. Technically, the
 formulation used in Refs.~\cite{fagotti14tdep,bertini15a} appears to be different from ours.

In the following, we will first introduce our model, a Heisenberg model with small perturbations described by Lindblad operators. Then we derive equations of motion for a time-dependent GGE and similar equations for block-diagonal density matrices. We then compare three types of approaches: the exact evolution of the density matrix, an approximate time-evolution in the subspace of block-diagonal density matrices and the evolution described by truncated GGEs.


\section{Model}\label{SecModel}
We consider the one-dimensinal spin-$1/2$ Heisenberg model
\begin{align}
H_0= J \sum_j  \boldsymbol S_j  \cdot \boldsymbol S_{j+1} 
\end{align}
arguably the most studied integrable system.  In  \cite{lange17} we investigated the case where a Heisenberg model was coupled to Hamiltonian perturbations arising from phonons and oscillating fields. This situation was, however, too complicated for a detailed numerical study on the validity of GGEs. Therefore we consider a numerically more tractable case and describe the (weak) integrability breaking by the coupling  to  non-equilibrium Markovian baths described by Lindblad dissipators $\cl{D}^{(i)}$ acting on the density matrix $\rho$ prepared at time $t=0$ in some initial $\rho(0)$
\begin{align}\label{d0}
\partial_t \rho &=(\cl{L}_0 +\cl{L}_1)  \, \rho \\
\mathcal{\hat{L}}_0 \rho &= -i[H_0,\rho], \
\cl{\mathcal{L}}_1\rho= \epsilon\left( \gamma \cl{D}^{(1)} + (1-\gamma) \cl{D}^{(2)} \right) \rho\nonumber
\end{align}
with
\begin{align}
 \cl{D}^{(i)}&=J \sum_{k} \left( L^{(i)}_k \rho {L^{(i)}_k}^{\dagger}- \frac{1}{2} \{{L^{(i)}_k}^\dagger L^{(i)}_k,\rho\} \right)
\end{align}
where $L_k$ are so-called Lindblad operators and the prefactor $J$ has been included to obtain a dimensionless $\epsilon$.  As Lindblad operators we chose
\begin{align}\label{EqLindblad}
L_k^{(1)}=S_k^z \quad \text{and} \quad  L_k^{(2)}=\frac{1}{2} (S_k^+ S_{k+1}^- + i S_{k+1}^- S_{k+2}^+)
\end{align}
$L_k^{(1)}$ represents dephasing and, considered alone, heats up the system to an infinite temperature state. $L_k^{(2)}$ has been chosen to break all relevant symmetries (up to $S^z$ conservation) as we want to study below the generation of heat currents. It also provides a cooling mechanism. We have checked that similar agreement is also obtained for other Lindblad operators. Our analysis will be performed as a function of the relative strength of both terms, $\gamma \in [0,1]$.
Using the conservation of the total magnetization $S^z$, we study in the following only the sector with $S^z=0$.

\section{Time-dependent generalized Gibbs ensembles}\label{SecFullDin}

The main goal of our paper is to provide numerical evidence for the following claim: 
The time evolution of a translationally-invariant integrable many-particle system in the presence of weak integrability-breaking perturbation is generically described by a generalized Gibbs ensemble
\begin{eqnarray}\label{dynamicalGGE}
\lim_{\epsilon \to 0} \rho(t)&\stackrel{\rm loc}{=}&\rho_{GGE}(t) \quad \text{for } t=\frac{1}{\epsilon J} \tau\\
 \rho_{GGE}(t)&=&\frac{e^{- \sum_i \lambda_i(t) C_i}}{\text{Tr}[e^{- \sum_i \lambda_i(t) C_i}]}
\end{eqnarray}
where $C_i$ are the (quasi-)local conservation laws of the integrable system \cite{grabowski96,ilievski16}.
Eq.~\eqref{dynamicalGGE} holds only in the thermodynamic limit and the $\stackrel{\rm loc}{=}$ symbol is used to indicate that it applies only to local observables $A$ for which
$\lim_{\epsilon \to 0} \tr[ A\, \rho(t)] = \tr[ A \, \rho_{GGE}(t)]$. We assume that at $t=0$ the dynamics is switched on and we have introduced the dimensionless time $\tau>0$ to indicate that the relation holds only for times of the order of $1/\epsilon$ and larger. The $\lambda_j(t)$ are determined from Eq.~\eqref{EqForces} derived below.  
For the validity of Eq.~\eqref{dynamicalGGE} we furthermore demand that Eq.~\eqref{EqForces} is well-behaved, leading to a unique steady state, see below.

In the limit $\tau \ll1 $ (with $\tau/\epsilon \gg 1$ such that $t \gg 1/J$), the effect of the perturbation can be ignored and one recovers the standard quench problem for an integrable system, for which the emergence of a GGE has been firmly established, see e.g. \cite{ilievski15a}. The initial values of the $\ave{C_i} = \tr[C_i \, \rho(0)]$ determine the value of the initial Lagrange parameters $\lambda_j(\tau \to 0)$. This regime is closely associated to the so-called prethermalization plateau,
where approximate conservation laws fix a transient state before perturbations set in, see e.g. \cite{bertini15}. The dynamics for $\tau >0$ is the focus of our study.

The dynamics of the Lagrange parameters is obtained by demanding that 
\begin{align}\label{CnGGE}
\tr[ C_i \,\dot \rho(t)]\stackrel{!}{=} \tr[ C_i \,\dot \rho_{GGE}(t)]
\end{align}
up to corrections which vanish for small $\epsilon$.
Approximating on the left-hand side of the equation $\dot \rho=(\cl{L}_0+\cl{L}_1) \rho \approx (\cl{L}_0+\cl{L}_1) \rho_{GGE}=\cl{L}_1 \rho_{GGE}$ and using $\dot \rho_{GGE}(t)=-\sum \dot \lambda_j  \rho_{GGE}(t)\,(C_j-\langle C_j\rangle_{GGE})$ with $\ave{A}_{GGE}=\tr[A \, \rho_{GGE}(t)]$ on the right-hand side, one obtains a simple  differential equation
\begin{align}\label{EqForces}
\dot{\lambda}_i&=F_i(t)  
\end{align}
where the generalized forces $F_i(t)$ are functions of $\lambda_j(t)$ obtained from 
\begin{align}
F_i(t)&\approx - \sum_j (\chi(t)^{-1})_{ij} \text{Tr}[C_j \cl{L}_1 \rho_{GGE}(t)] \\
&=-\sum_j (\chi^{-1})_{ij} \langle \dot{C}_j \rangle_{GGE} \nonumber \\
\chi_{ij}(t)&= \langle C_i C_j \rangle_{GGE}- \langle C_i\rangle_{GGE} \langle C_j \rangle_{GGE} \notag
\end{align}
Note that the forces $F_i$ are of order $\epsilon$ (correction to Eq. (\ref{EqForces}) are of order $\epsilon^2)$. Therefore the time evolution after the initial prethermalization is slow and set by a time scale of order $1/\epsilon$. More precisely, this is valid for perturbations of the  Lindblad type studied in this paper. For Hamiltonian perturbations, the linear order perturbation theory vanishes. As is well-known from Fermi's golden rule,  transition rates arise only in second order perturbation theory. The formulas given above can easily be generalized to this case, see Methods section of Ref.~\cite{lange17}.

An alternative, slightly more formal derivation of Eq.~(\ref{EqForces}) is obtained by setting
$\rho(t)=\rho_{GGE}(t)+\delta \rho(t)$ where $\rho_{GGE}(t)$ has to be chosen in such a way that $\delta \rho \sim \epsilon$ vanishes in the limit $\epsilon \to 0$. The essential idea is now to separate the slow dynamics within the GGE manifold arising from $\cl{L}_1 \sim \epsilon$ from the fast dynamics in the perpendicular space. One therefore introduces a projection operator \cite{lenarcic18,Robertson66,Robertson68,kawasaki73,Koide02,breuer02} 
\begin{align}
\hat{P}(t) X:=- \sum_{i,j} \frac{\partial \rho_{GGE}(t)}{\partial \lambda_i} (\chi(t)^{-1})_{ij} \text{Tr}[C_j X]
\end{align}
which projects density matrices onto the space tangential to the GGE manifold, spanned by $\partial \rho_{GGE}(t)/\partial \lambda_i$ . We apply $\hat P(t)$ to Eq.~(\ref{d0}) and use that $\hat P(t) \dot \rho_{GGE}(t)=\dot \rho_{GGE}(t)$,
$\cl{L}_0 \rho_{GGE}=0$,  $\hat P(t) \cl{L}_0 \delta \rho=0$, and $\cl{L}_1 \delta \rho \sim \epsilon^2$.
From this we  obtain
\begin{align}
\hat P(t) \delta \dot{\rho} + \dot \rho_{GGE} = \hat P(t) \cl L_1 \rho_{GGE} + \mathcal O(\epsilon^2)
\end{align}
Demanding that $\hat P(t) \delta \dot{\rho}$ is of order $\epsilon^2$ leads to  $\dot \rho_{GGE} = \hat P(t) \cl L_1 \rho_{GGE} $ which is equivalent to  Eq.~(\ref{EqForces}).

The arguments given above, strongly suggest that the GGE ansatz fulfills the time evolution equation~\eqref{d0} projected on the conservation laws
up to corrections of order $\epsilon^2$. This does, however, not yet guarantee the validity of the much stronger claim that the time-dependent GGE is also valid in the long-time limit. For this we have to demand that errors don't pile up during time evolution but decay exponentially. This is the case in situations where Eq.~(\ref{EqForces}) predicts a unique and stable steady state, see Appendix \ref{AppendixForces}. In all examples considered by us so far, we have never found that this condition is violated. 

We will show that in practical implementations it is not necessary to take all $O(N)$ (quasi-)local conservation laws into account. Accurate results can already be obtained for a truncated GGE (tGGE), including only a small number of approximately conserved quantities and Lagrange parameters.

\section{Time-dependent block-diagonal density matrices}\label{SecTimeDiag}
The GGE approach is only valid in the thermodynamic limit where the (quasi-)local approximately conserved $C_i$ determine the dynamics. For smaller systems, one has, however, to take into account that the set of conservation laws of $H_0$ is much larger, and given by $\mathcal{Q}=\{ \ket{n} \bra{m} \ \text{with} \ E_n^0=E_m^0 \}$ where $\ket{n}$ and  $\ket{m}$ are eigenstates of $H_0$ with the same energy.
Note that the elements of $\mathcal{Q}$ are in general non-commuting and highly non-local operators.  In the limit of small $\epsilon$, one can, however, derive the dynamics in the space spanned by the elements of $\mathcal{Q}$. Such approaches are well described in literature \cite{breuer02} and we briefly sketch the relevant formulas, emphasizing the analogy to the GGE approach.
 
The role of the GGE density matrix  is taken over by the block-diagonal density matrix
\begin{align}\label{BD}
\rho_{BD}(t)=\sum_{E_n=E_m} \lambda_{nm}(t)\,  \ket{n} \bra{m} 
\end{align}
with normalization $\sum_n \lambda_{nn}=1$. 
In analogy to Eq.~(\ref{CnGGE}), we demand that $\tr[\ket{m} \bra{n}\,  \dot \rho(t)]\stackrel{!}{=} \tr[\ket{m} \bra{n} \, \dot \rho_{BD}(t)]$ up to corrections vanishing for $\epsilon \to 0$. Using $\dot \rho(t) \approx \cl{L}_1 \rho_{BD}(t)$, we obtain a linear (!) differential equation for  $\lambda_{nm}(t)$ 
\begin{align}\label{BDtime}
\dot \lambda_{nm}(t)=\sum_{E_{n'}=E_{m'} } M_{nm,n'm'}   \,                        \lambda_{n'm'}(t)  
\end{align}
where $M$ is an effective Liouvillian acting on the space of block-diagonal density matrices ($E_n=E_m$, $E_{n'}=E_{m'}$)
\begin{align}
M_{nm,n'm'}=\tr[\,\ket{m} \bra{n} \, \cl{L}_1 \, \ket{n'} \bra{m'}\,]
\end{align}
$M$ is linear in $\epsilon$ and therefore all $\epsilon$ dependence can be absorbed in a rescaling of the time axis within this approximation which is valid only for small $\epsilon$ and covers the dynamics after prethermalization.
The initial condition for the time evolution is simply set by $\lambda_{nm}(0)=\bra{n} \rho(0) \ket{m}$. 

Compared to the GGE approach which uses only a few (maximally $O(N)$) Lagrange parameters, the block-diagonal matrix uses $O(2^N)$ parameters and is therefore much less efficient. For $N=14$ and $S^z=0$ we have to use 
6752 parameters.
The block-diagonal approach is, however, numerically much more efficient than an approach using the full density matrix which has $O(4^N)$ parameters.

\section{Numerical Results }\label{SecResults}
To test the validity of the GGE approach we have to face the problem that the GGE approach is only valid in the thermodynamic limit while the exact results for driven nonequilibrium systems at finite $\epsilon$ can only be obtained for tiny systems.
We therefore use the following two-step approach: We first show numerically that for small systems ($N=8$) the numerically exact results obtained from the exact time-dependent density matrices for small $\epsilon$ are well-described by time-dependent block-diagonal density matrices. We then compare for larger systems (up to $N=14$) the block-diagonal density matrices to truncated GGEs based on only a small number of approximately conserved quantities.

\begin{figure}[h!]
\center
\includegraphics[width=.99\linewidth]{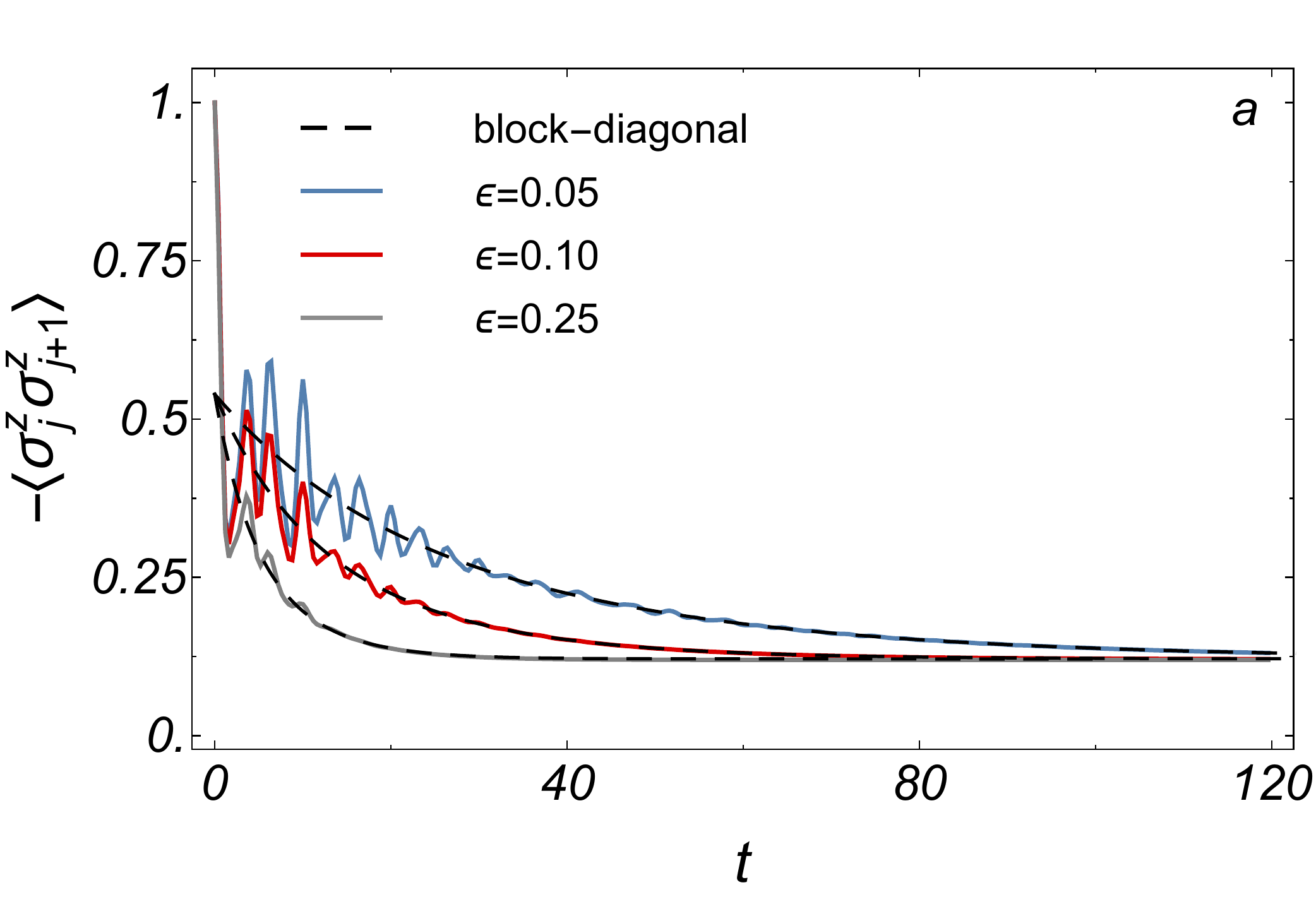}\\
\includegraphics[width=.99\linewidth]{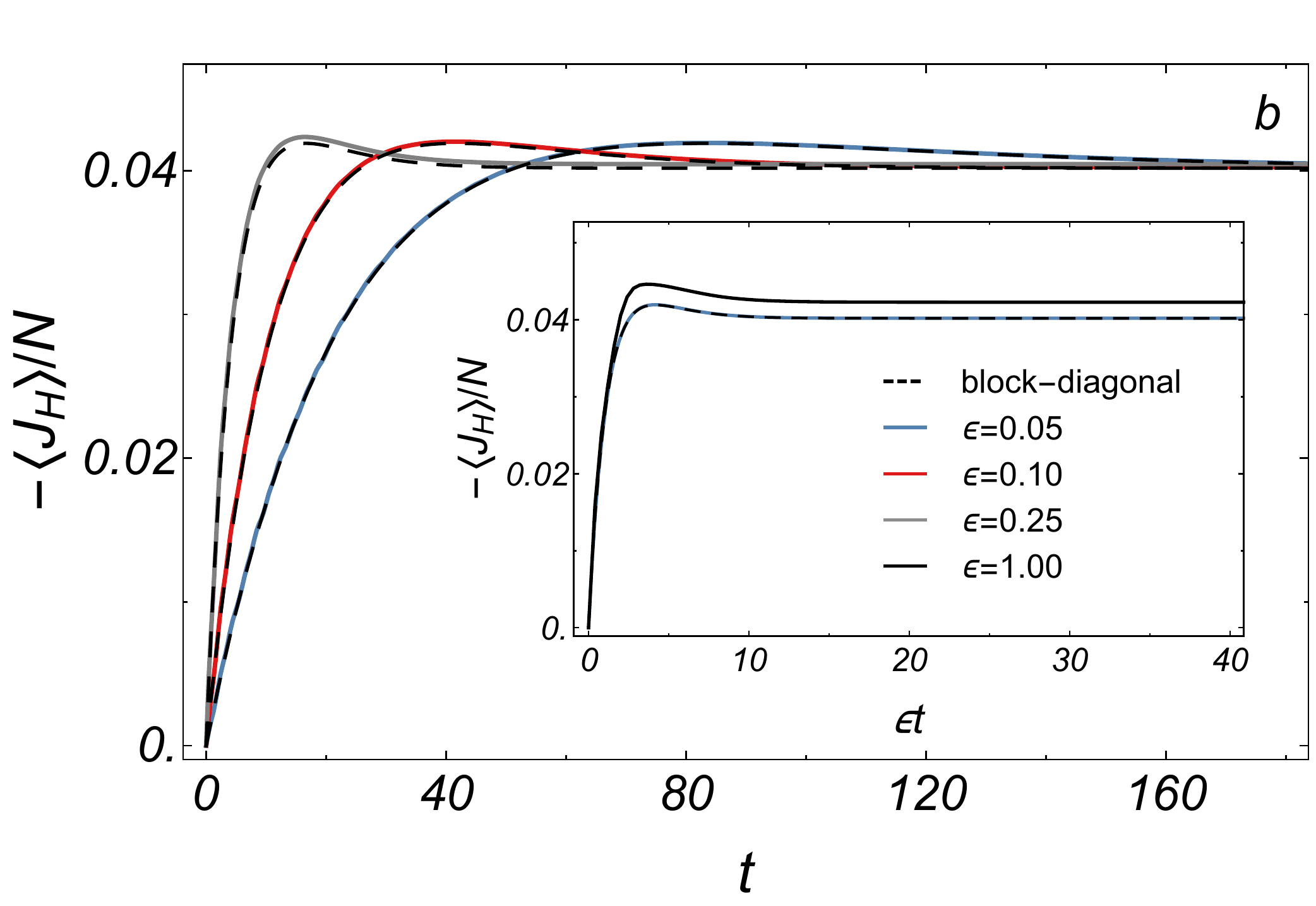}
\caption{\label{FigTimeExact} Exact time evolution of a weakly perturbed Heisenberg model ($N=8$, $\gamma=0.8$, $J=1$) for 
three small values of $\epsilon$. The dashed lines show the result for the time evolution of the block-diagonal density matrix using Eqs.~(\ref{BD},\ref{BDtime}). (a) Decay of the nearest-neighbor spin-correlation. (b) For the heat current rapid oscillations on a time scale of order $1/J=1$ are absent as $J_H$ is a conservation law of $H_0$. For the values of $\epsilon$ shown in the plot the dashed lines follow the solid lines: the block-diagonal density matrices correctly describe the time evolution with high precision. Inset: At large $\epsilon=1.0$ discrepancies are visible.}
\end{figure}

\subsection{Time evolution for small systems}
As an initial state, we consider a classical N\'eel configuration. In Fig. \ref{FigTimeExact} we show the time evolution of the nearest-neighbor spin correlation, $\ave{\sigma^z_i \sigma^z_{i+1}}$, and of the heat current, 
\begin{align}
J_H= J^2 \sum_j (\boldsymbol S_j \times \boldsymbol S_{j+1})\cdot \boldsymbol S_{j+2}
\end{align}  
for a small system with $N=8$ sites. We compare the numerically exact results, calculated from the exact density matrix evolution, to the approximate results based on the block-diagonal density matrix.
Starting from $1$  at $t=0$ the spin correlation rapidly decays on a time scale of order $1/J=1$ to a value around $0.45$. Due to the smallness of the system, rapid oscillations persist and get only slowly damped.  Subsequently both the average value of the spin-correlation and the oscillations decay on a time scale set by $1/\epsilon$. The block-diagonal density matrix correctly captures the decay of the spin-correlations quantitatively. The time-dependence of the heat current, in contrast, is much smoother as the heat current is a conserved quantity, $[J_H,H_0]=0$. The initial state has not heat current but a large heat current builds up on a time scale set by $1/\epsilon$. The heat current obtained in the long-time limit is large and the system is therefore far out of equilibrium. Its value is approximately independent of $\epsilon$ for small $\epsilon$ and is predicted by the time-dependent block-diagonal density matrices and the GGE approach, see below.
The main result of this section is, however, that  
for small $\epsilon$ the time evolution for times large compared to $1/J$ is  accurately described by the block-diagonal ensemble.

\begin{figure}[b!]
\center
\includegraphics[width=.99\linewidth]{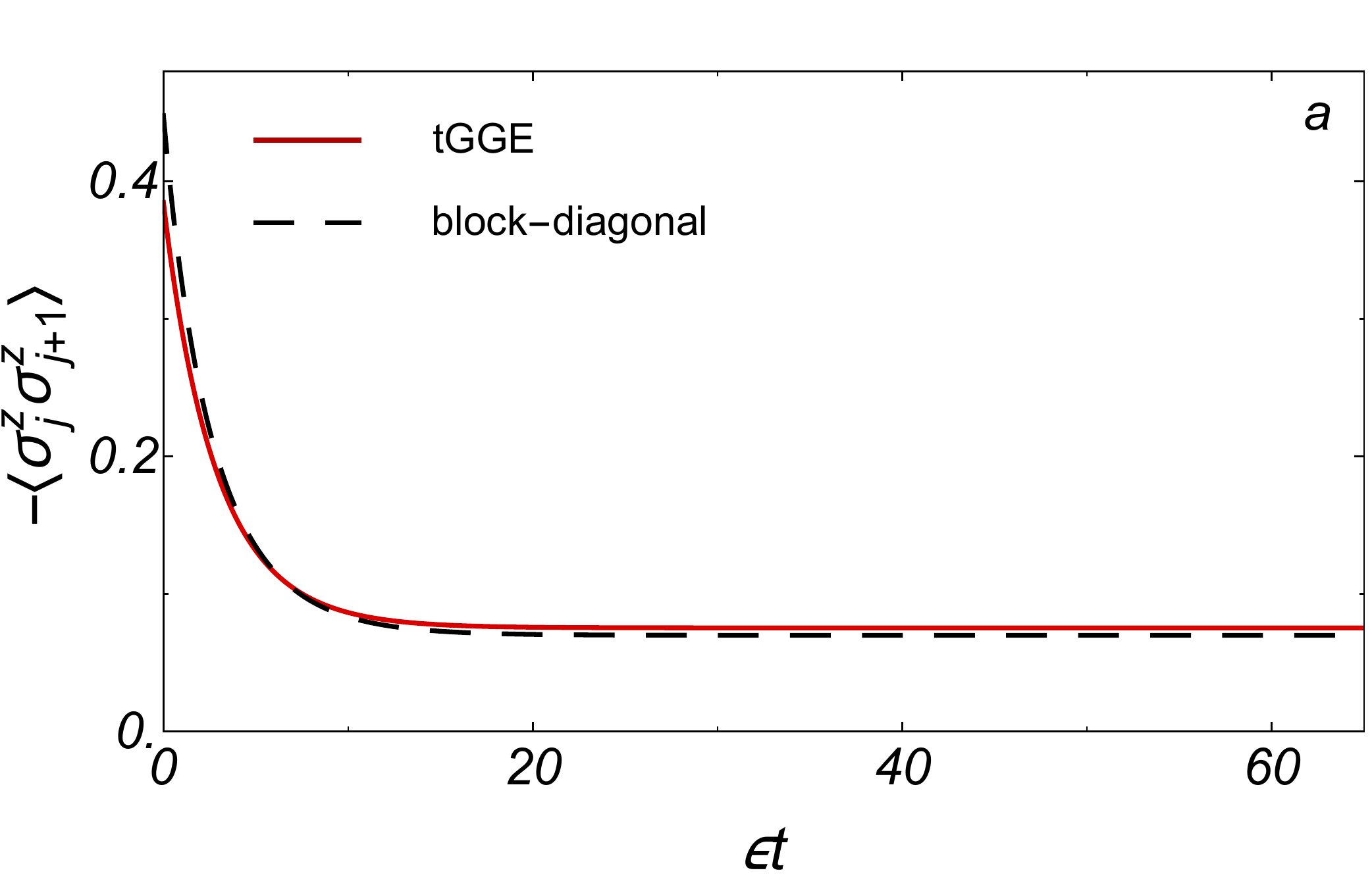}
\includegraphics[width=.99\linewidth]{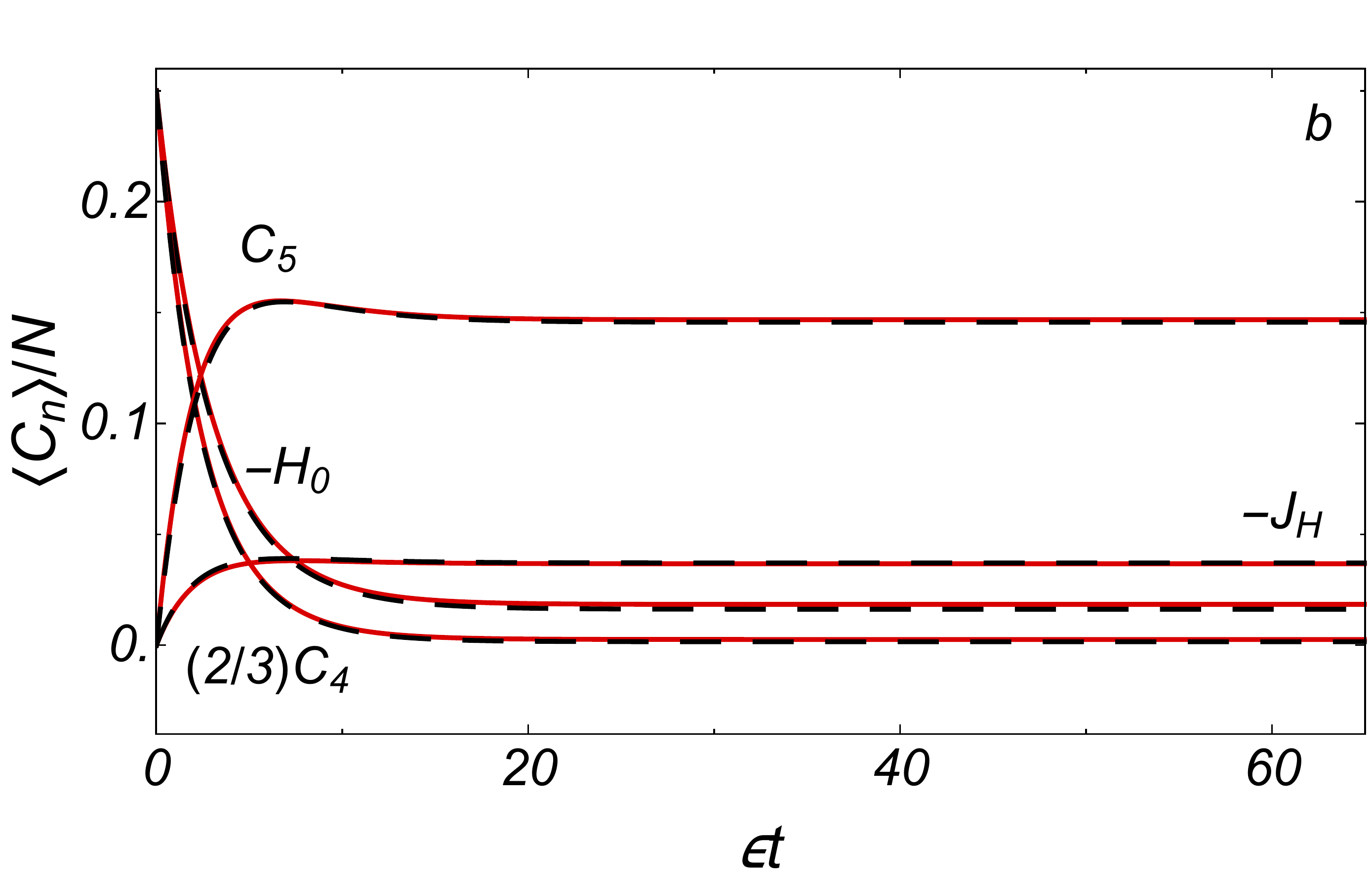}
\caption{ Comparision of the results obtained from the truncated GGE (solid line) and the block-diagonal density matrix (dashed). (a) Antiferromagnetic nearest neighbor spin correlations, (b) Approximate conservation laws, $C_5$, $-J_H$, $-H_0$, and $(2/3)C_4$. All quantities change on a time scale of order $1/\epsilon$. Parameters: $\gamma=0.8$, $J=1$, $N=14$.
\label{FigTimeGGE}}
\end{figure} 

\subsection{Time evolution of truncated GGE}
For a system with $N=14$ sites, we compare in Fig. \ref{FigTimeGGE} the time evolution of the block-diagonal ensemble with the results obtained for a truncated GGE based on only $N_c=4$ conserved quantities, $C_2, \dots, C_5$, where $C_2=H_0$ is the Hamiltonian, $C_3=J_H$ is the heat current and $C_4=[O_b,C_3]$, $C_5=[O_b,C_4]$ are conservation laws involving products of 4 and 5 spins. Here $O_b=-i \sum_j j \boldsymbol S_j  \cdot \boldsymbol S_{j+1}$ is the so-called boost operator
 \cite{grabowski96}. $C_1=S^z$, the total spin in z-direction, does not play a role in our study which focuses on the $S^z=0$ sector. The heat current operator $C_3$ and $C_5$ have the same symmetry properties.
Despite of the rather small system size, the small number of conservation laws and the omission of quasi-local conservation laws \cite{ilievski15,ilievski16}, a surprisingly accurate description of the time evolution is obtained. Note that the block-diagonal matrix approach keeps track of 6752 approximately conserved quantities to be compared to just 4 approximately conserved quantities in the truncated GGE!

The largest discrepancies are visible for the spin-spin correlation function at $t=0$. 
Note that this limit is completely independent of the integrability breaking perturbations. It only tests whether diagonal ensemble and GGE coincide after a quantum quench of the pure Heisenberg model. It therefore tests the ability of the GGE to describe the steady state after a quantum quench in an integrable system. Many previous numerical and analytical studies have shown that for this problem the GGE approach applies,  
e.g. \cite{ilievski15a,wouters14,piroli16gge}. 

\begin{figure}[b!]
\center
\includegraphics[width=.99\linewidth]{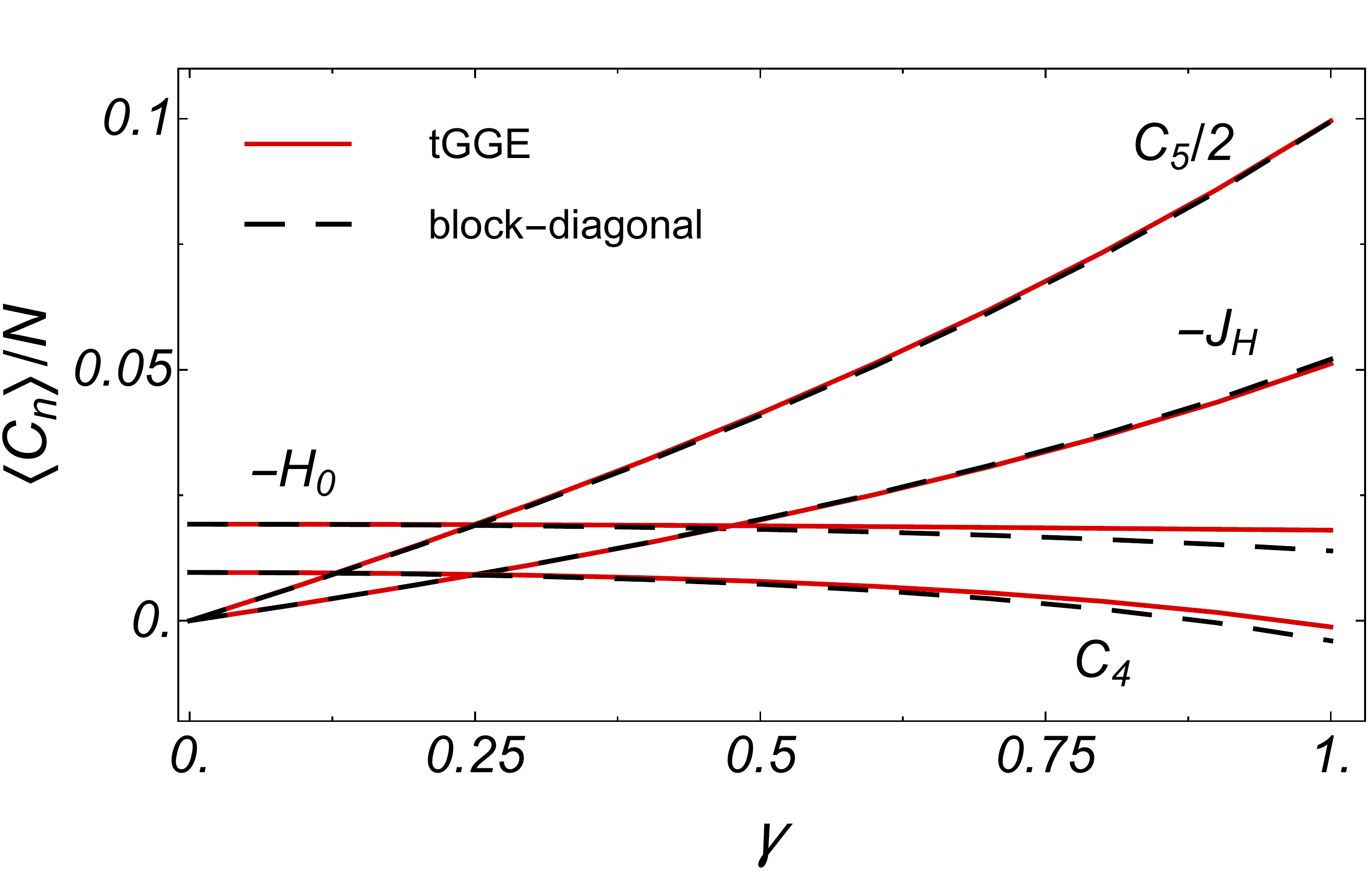}
\caption{Expectation values of conservation laws in the stationary state ($t\to \infty$) as a function of $\gamma$, which controls the nature of the dissipative terms ($N=14$, $J=1$). As in Fig. \ref{FigTimeGGE}, a comparison of the truncated GGE (solid lines) the block-diagonal density matrix (dashed) is shown. For $\gamma=0$ the system heats up to infinite temperatue and all conservation laws vanish in the thermodynamic limit. Due to finite size effects small finite values are obtained form $H$ and $C_4$.\label{figCnGamma}}
\end{figure} 
The lower panel of Fig. \ref{FigTimeGGE} shows the time evolution of the conserved quantities. In this case by construction the value at $t=0$ are the same for the truncated GGE and the block-diagonal approach. All conserved quantities change on time scales of order $1/\epsilon$ compared to their prethermalized value and show an exponential decay towards their steady-state value.
Note that features like the small overshooting of $C_5$ at intermediate times are well reproduced by the numerical approach.

The agreement obtained for $\gamma=0.8$ in Fig.  \ref{FigTimeGGE} is also observed for other values of the parameter $\gamma$ controlling the nature of the dissipative terms. This is shown in Fig.~\ref{figCnGamma}, where the steady-state value of the conservation laws is shown as function of $\gamma$. The largest deviations are visible for $\gamma$ close to $1$.

\subsection{Finite-size, finite-$\epsilon$, and truncation effects}

Formally, the description of the driven many-particle quantum system by a time-dependent GGE is only accurate in the limit of weak perturbations, $\epsilon \to 0$, for large systems, $N \to \infty$, and taken all (quasi-)local conservation laws into account, $N_c \to \infty$. The results presented above already suggest that one can, nevertheless, obtain surprisingly accurate results for moderate values of $\epsilon$, rather small system sizes and a tiny number of conservation laws.

\begin{figure}[b!]
\center
\includegraphics[width=0.99\linewidth]{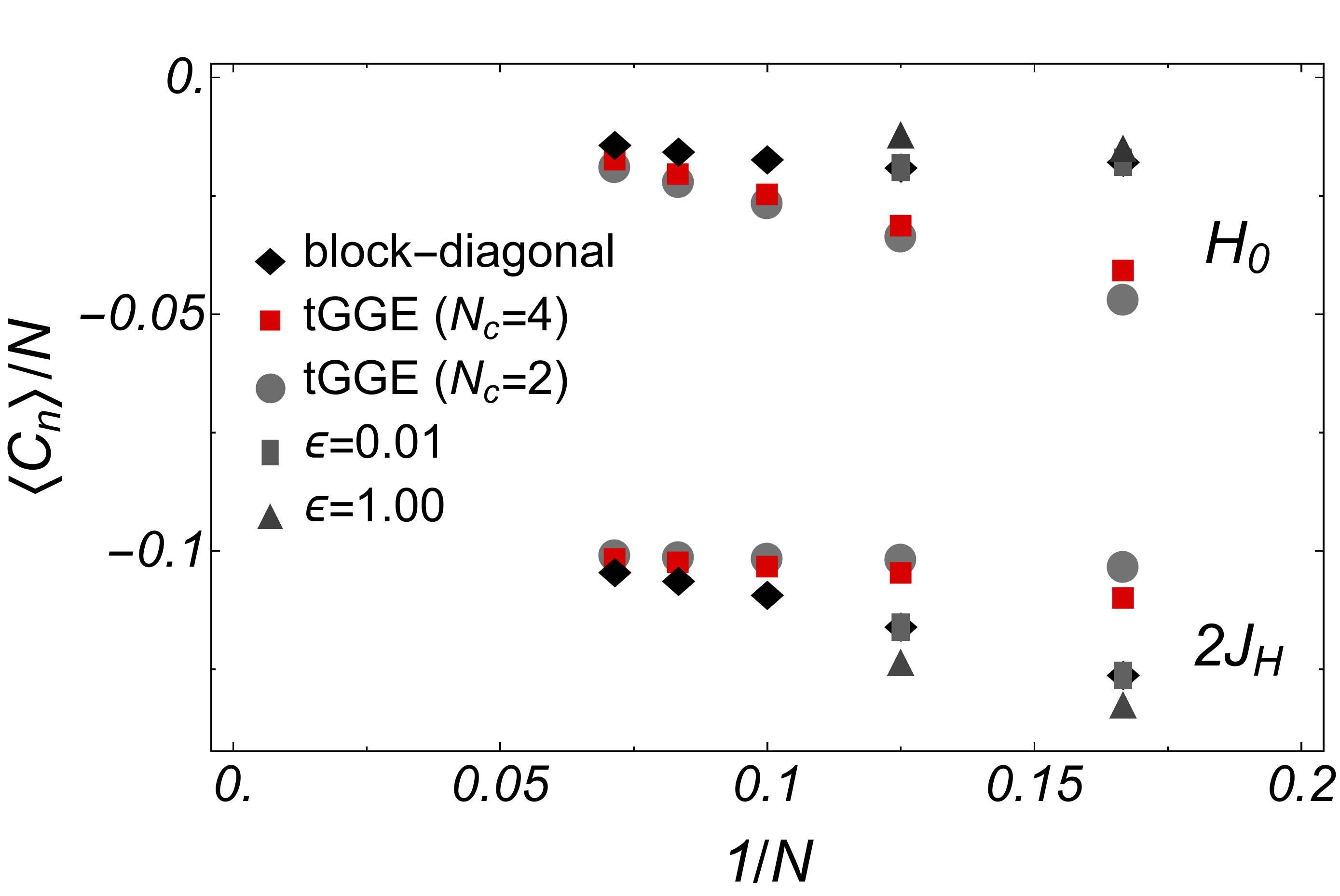}
\caption{\label{FigFiniteSize} Steady-state expectation value of the energy, $\ave{H_0}$, and the heat current, $\ave{J_H}$, as function of the inverse system size, $1/N$ ($N=6,8,10,12,14)$, for the truncated GGE with 2 and 4 conservation laws and the block-diagonal ensemble. For the smallest system size, $N=6,8$, we also show the exact results for $\epsilon=0.01$, which practially coincides with the the block-diagonal ensemble, and for $\epsilon=1.0$ ($\gamma=1$).}
\end{figure} 

In Fig.~\ref{FigFiniteSize} the expectation value of the energy density and the heat current in the steady state ($t \to \infty$) are shown as function of $1/N$ for the exact density matrix, for the block-diagonal ensemble (exact for $\epsilon \to 0$) and for two truncated GGEs with $N_c=2$ and $N_c=4$. One clearly sees that in the thermodynamic limit the truncated GGEs become more and more accurate. Already for the largest system ($N=14$) for which we were able to evaluate the block-diagonal ensemble, a satisfactory agreement is obtained. Also $N_c=4$ is more accurate than $N_c=2$ but the errors arising from finite-size effects are dominating. We are therefore not showing results for larger values of $N_c$ as those are spoiled by finite-size effects which tend to become more severe for more complicated approximate conservation laws which involve a large number of neighboring spins.

\begin{figure}[t!]
\center
\includegraphics[width=\linewidth]{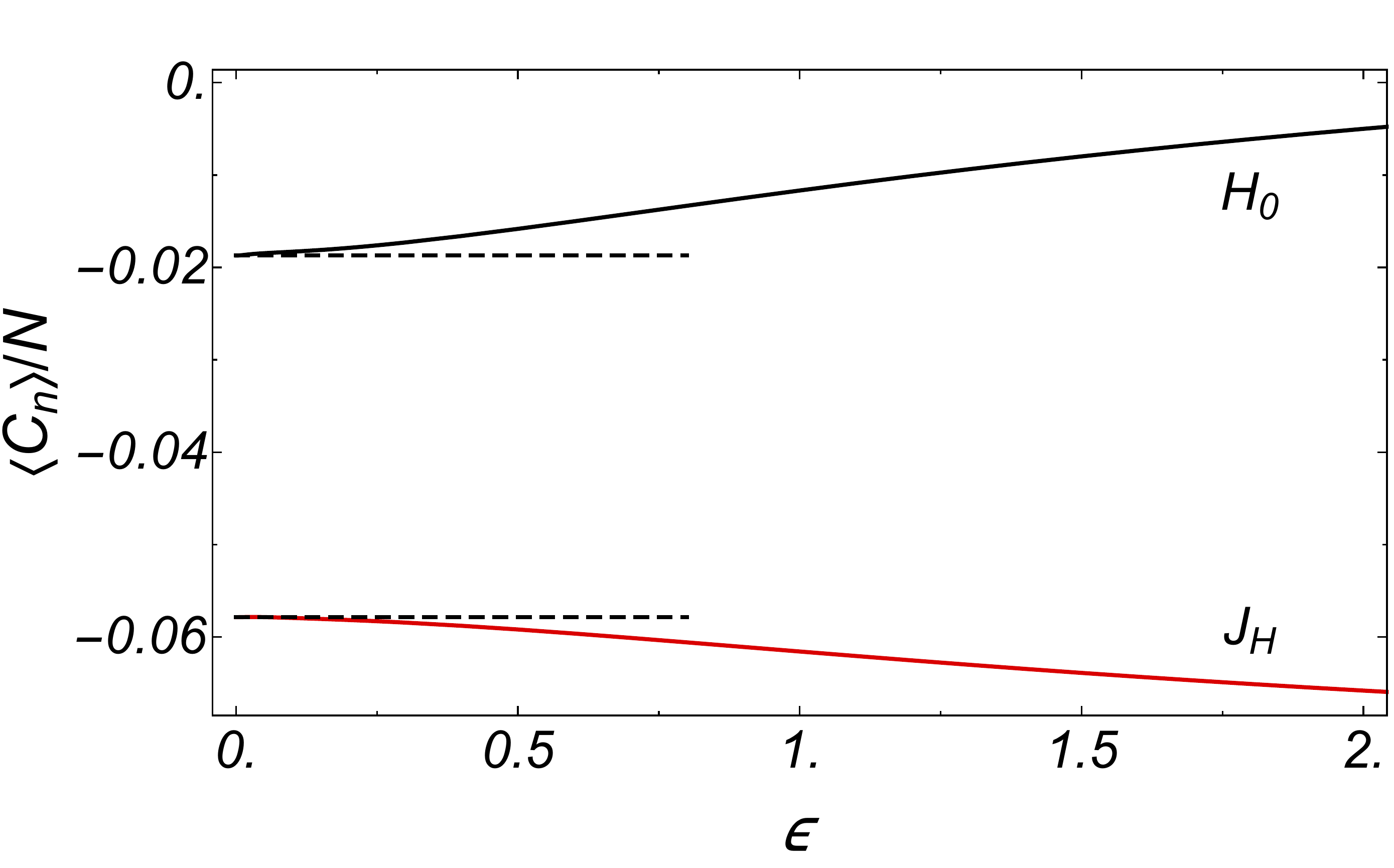}
\caption{\label{FigFiniteEps}  Steady-state expectation value of the energy, $\ave{H_0}$, and the heat current, $\ave{J_H}$ as function of the strength $\epsilon$ of the integrability-breaking Lindblad terms ($N=8$, $J=1$, $\gamma=1$). For $\epsilon \to 0$ the result of the block-diagonal ensemble (dashed line) is recovered.}
\end{figure}

The effects of finite $\epsilon$ for steady-state expectation values are displayed in Fig.~\ref{FigFiniteEps} for $N=8$. We find that the $\epsilon$ dependence of the steady state is not strongly pronounced and is described by a smooth function. In the $\epsilon \to 0$ limit the block-diagonal ensemble is exact (as expected from the analytical arguments). It also captures with high accuracy the properties for small $\epsilon$.

The results obtained in the limit $\epsilon \to 0$ can be systematically improved using perturbation theory in $\epsilon$, developed for the steady state in Ref.~\cite{lenarcic18}. Here it is important to distinguish the perturbation theory for finite size systems ($\epsilon$ smaller than the dimensionless level-spacing $1/N$ or even $2^{-N}$) from the perturbation theory in the thermodynamic limit  ($ 1/N \ll \epsilon \ll 1$). The numerical results show indeed a different behavior in the regime $\epsilon \lesssim 0.1$ and $0.1\lesssim \epsilon \lesssim 1$ but the system size is too small to extract reliable results for the perturbation theory in the thermodynamic limit. A more detailed discussion of this issue can be found in the Appendix \ref{AppPT}.

\vspace{.5cm}

\section{Conclusion and Outlook}

We have demonstrated that time-dependent generalized Gibbs ensembles can be used to describe quantitatively the dynamics of approximately integrable systems where small perturbations drive the system far from equilibrium.
While the present study has focused on perturbations arising from Lindblad operators describing the coupling to Markovian baths, it can also be used to investigate Hamiltonian perturbations. Our example, a Heisenberg model coupled to two types of Lindblad disspators was chosen for numerical convenience but one can think of a wide range of experimental systems where our approach is applicable. In practically all cold-atom experiments there are atomic loss processes. An interesting question is therefore how atomic loss processes affect experimental ultracold-atom realization integrable models, e.g., of the fermionic Hubbard model in one dimension. It is reasonable to assume that the loss processes will activate some of the exotic conservation laws of this model. 
An experimental setup, particularly suitable for our theoretical proposal, is also that of trapped ions where openness can be directly simulated \cite{barreiro11} by realizing Lindblad driving \cite{Reiter12}, currently using a few tens of atoms \cite{friis17}.
Another interesting class of systems are spin-chain materials, well-described by one-dimensional Heisenberg models. Here phonons and the coupling to lasers take over the role of the integrability breaking perturbations.  We have studied steady-state properties of such models in Ref.~\cite{lange17}.

We have used exact diagonalization of Lindblad operators to be able to compare the time-dependent GGE approach to exact results. A main advantage of the time-dependent GGE approach is that it can be combined with other, more powerful numerical approaches. For  Markovian dynamics it is sufficient to evaluate simple  expectation values of operators to calculate effective forces. Many different numerical or analytical methods can therefore be used to obtain the non-equlibrium dynamics. This includes Monte-Carlo approaches, transfer-matrix DMRG methods, or high-temperature expansions. For the model considered by us all Lagrange parameters remain rather small during time evolution. Therefore it should be possible to calculate the dynamics of a truncated GGE using a rather straightforward high-temperature expansion (or, more precisely, small-Lagrange-parameter expansion)  directly in the thermodynamic limit, $N=\infty$. 

There are many interesting open question. For example, for the Lindblad driving studied by us the solution of the rate equation for Lagrange parameters, Eq.~(\ref{EqForces}), shows a simple exponential relaxation to a single steady state.
Out of equilibrium, however, other types of behavior can also occur: several steady states, cyclic solutions, or even chaotic solutions. It is an interesting open question how our approach has to be modified in these cases. Another interesting class or problems concern situations which are not translationally invariant and where Lagrange parameters depend on space and time. For exactly integrable models such a hydrodynamics description has recently be developed \cite{bertini16a,castro-alvaredo16,bulchandani17}, and we expect that it can be generalized in a straightforward way to models where integrability is broken by small perturbations
which drive the system out of equilibrium.

\section*{Acknowledgement}
We acknowledge useful discussions with B. Bertini and financial support of the German Science Foundation under CRC 1238 (project C04) and CRC TR 183 (project A01).


\appendix
\section{Perturbation theory for steady state}\label{AppPT}
We have argued that in the limit of small but finite perturbation
strength $\epsilon$ a time-dependent GGE or an approach based on
block-diagonal density matrices correctly describes both the dynamics
and the steady state of the perturbed integrable model.

\begin{figure}[b!]
\center
\includegraphics[width=0.485 \linewidth]{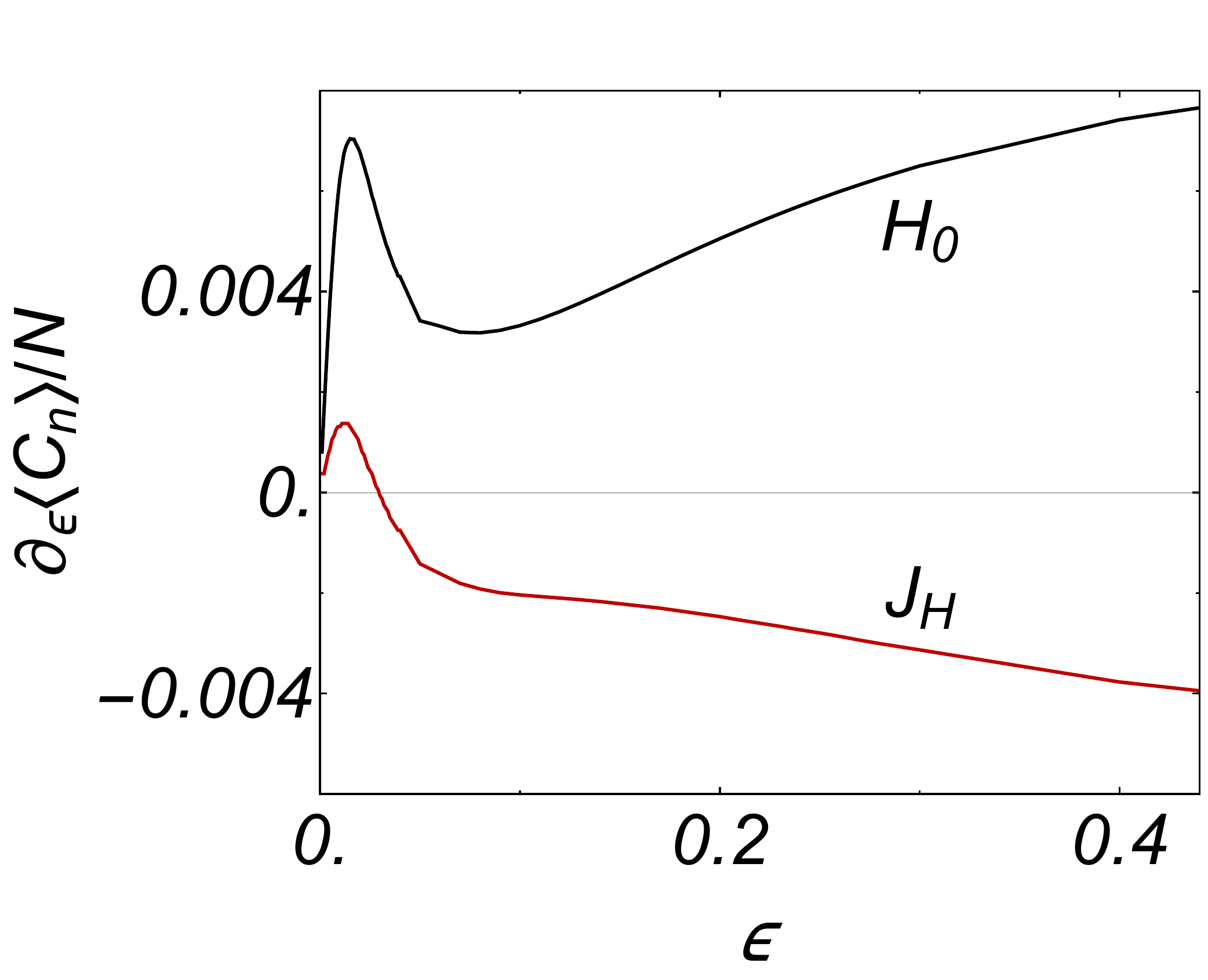}
\includegraphics[width=0.495 \linewidth]{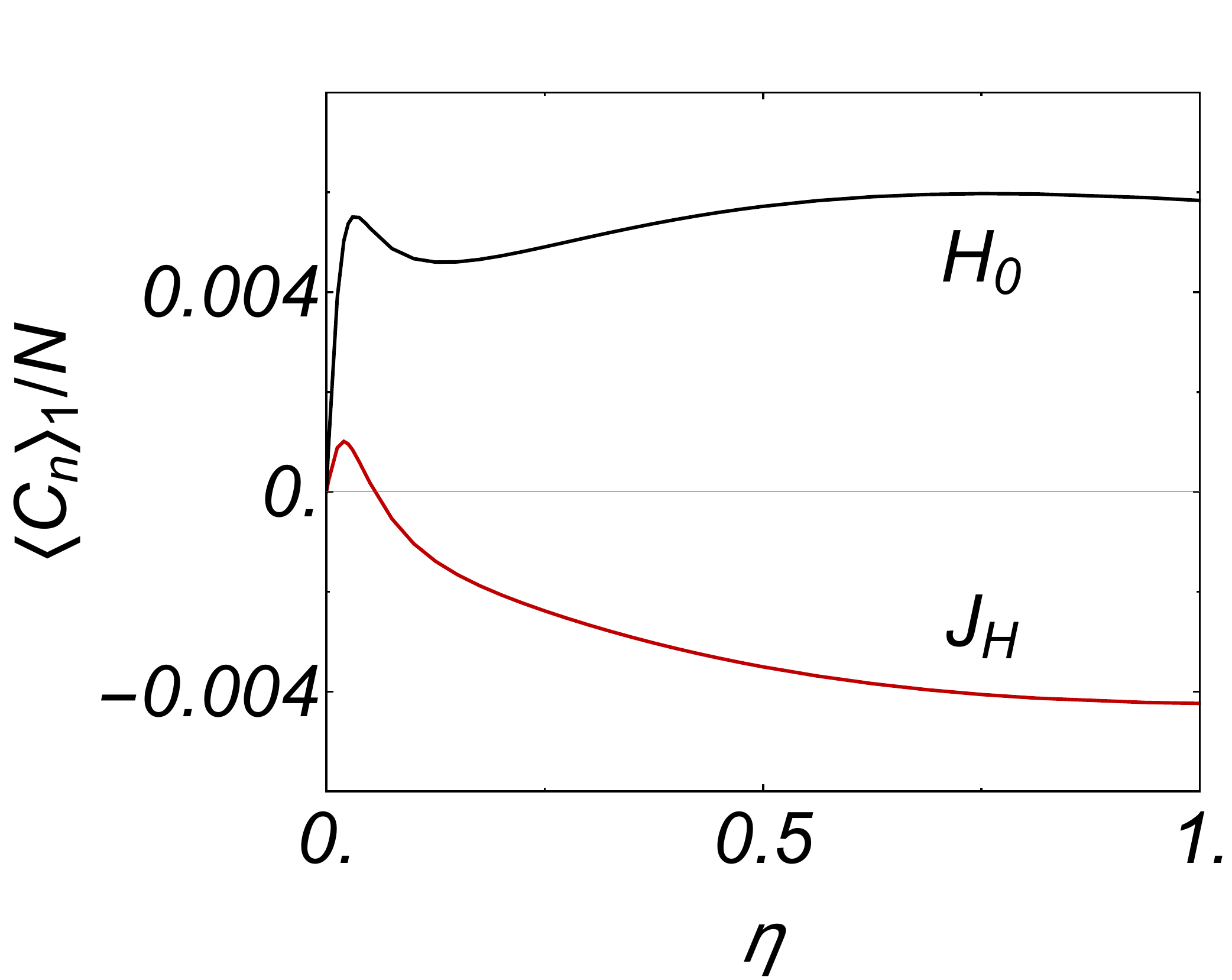}
\caption{\label{figPT} Left panel: Derivative of energy and heat current, $\frac{d \ave{H}}{d \epsilon}$ and $\frac{d \ave{J_H}}{d
  \epsilon}$, as function of the strength of perturbation $\epsilon$ calculated for $\gamma=1$, $J=1$ and $N=8$. Right panel: Calculation of the coefficient linear in $\epsilon$ from perturbation theory \cite{lenarcic18} as function of the broadening $\eta$ for the same parameters.}
\end{figure}

In this Appendix we discuss the leading order correction to the steady state for finite $\epsilon$. 
In Ref.~\cite{lenarcic18} we have shown how one can formulate a perturbation theory in powers of $\epsilon$ around such a state. Here it is important to distinguish the perturbation theory in the thermodynamic limit from the perturbation theory for finite-size
systems. The latter is only valid for $\epsilon$ small compared to the level spacing $\Delta=\delta J$ of the system, where $\delta$ is the dimensionless level spacing. We are mainly interested in the opposite limit $\delta  \ll \epsilon \ll 1$. 

As is well known from standard perturbation theory (Kubo formula) it is essential to include a small imaginary decay rate $i \eta$ in all calculations. For $\eta \ll \Delta$ one recovers the perturbation theory for a finite-size system while in the thermodynamic limit one choses $\eta \gg \Delta$ but smaller than all other relevant energy
scales.

As our goal is to compare numerically exact results with the formulas of  Ref.~\cite{lenarcic18}, we have to face the problem that exact results are only available for rather small system sizes and 
therefore the regime $\delta \ll \epsilon \ll 1$ and $\Delta \ll \eta \ll J$ are difficult to achieve.

As we are interested in the correction {\em linear} in $\epsilon$, we plot in Fig.~\ref{figPT} (left panel) the derivative of energy and heat-current, $d \ave{H}/d \epsilon$ and $d \ave{J_H}/d\epsilon$. This is compared to the perturbation theory result to linear order in $\epsilon$ (based on the formulas derived in Ref.~\cite{lenarcic18}) shown in the right panel as function of the broadening $\eta$.  Both panels show that the linear slope vanishes in a finite size system. For finite $N$ and in the steady state the leading correction is of order $\epsilon^2/\delta$. In the analytic treatment this can be shown by observing that  the corrections to the steady state are proportional to the imaginary part
of $\frac{1}{E_n-E_m-i \eta}$ with $E_n \neq E_m$ \cite{lenarcic18} which vanishes for
$\eta \ll \Delta$.

In the thermodynamic limit, $1 \gg \epsilon \gg \delta$, we expect instead that a linear correction does exist. Unfortunately, we cannot extract a well-defined linear slope from the exact result shown in Fig.~\ref{figPT} (left panel) due to the small size of the system. 
The same issue arises also in the dependence of the perturbative result of $\eta$. The fact that qualitatively similar results are obtained for the $\eta$ and  $\epsilon$ dependencies is not an accident but reflects that the Lindblad coupling effectively leads to a broadening of levels. 

In conclusion, our analysis has shown that it is very important to distinguish perturbations for finite size systems and in the thermodynamics limit. At least semi-quantitatively, the analysis also confirms the perturbative approach suggested in Ref.~\cite{lenarcic18}.

\section{Effective forces and $\gamma$ dependence of steady-state expectation values}\label{AppendixForces}
\begin{figure}[h]
\center \includegraphics[width=\linewidth]{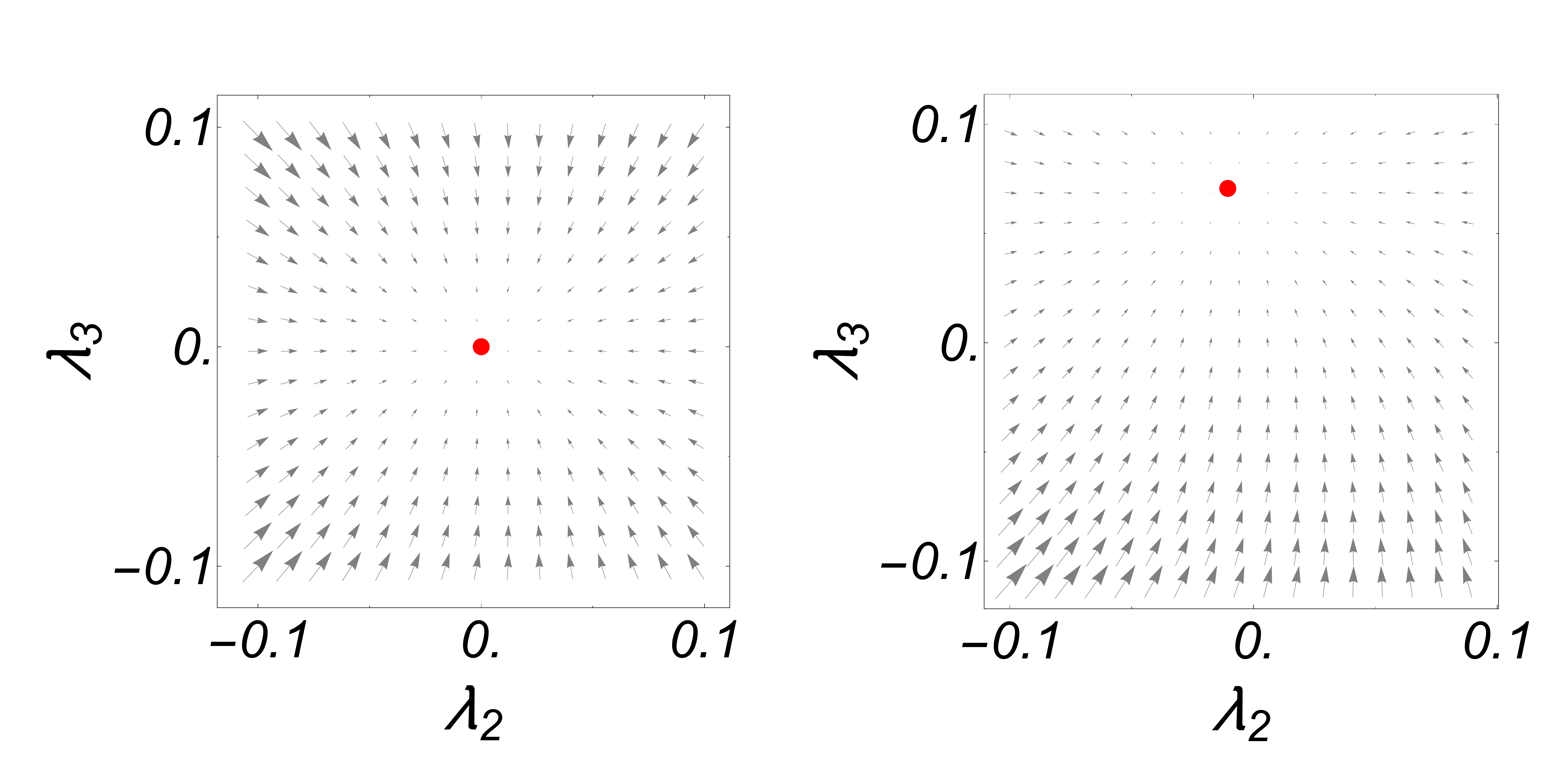}
\caption{Effective force field $(F_2,F_3)$ in the vicinity of the steady state (red point) in the plane spanned by the Lagrange parameters $\lambda_2=\beta$ and $\lambda_3$ using $e^{- \beta H-\lambda_3 J_H}$ as an ansatz for $\rho_{GGE}$ ($N=8$). Left: For $\gamma=0$ the system approaches an infinite temperature state since the Lindblad operator $L^{(1)}$ is constantly heating the system up. Right: At $\gamma=1$ the system is attracted towards a non-equilibrium state with finite stationary values of both $\lambda_2$ and $\lambda_3$. \label{figForces} }
\end{figure}

In Fig.~\eqref{figForces} we show the effective forces which determine the dynamics of the time-dependent GGE according to Eq.~\eqref{EqForces}. For the chosen Lindblad dynamics we find that the system is always attracted to a unique, well-defined fixed point. Therefore also small errors in, e.g., the initial state do not grow over time but are damped out.

\begin{figure}[h!] 
\includegraphics[width=.85\linewidth]{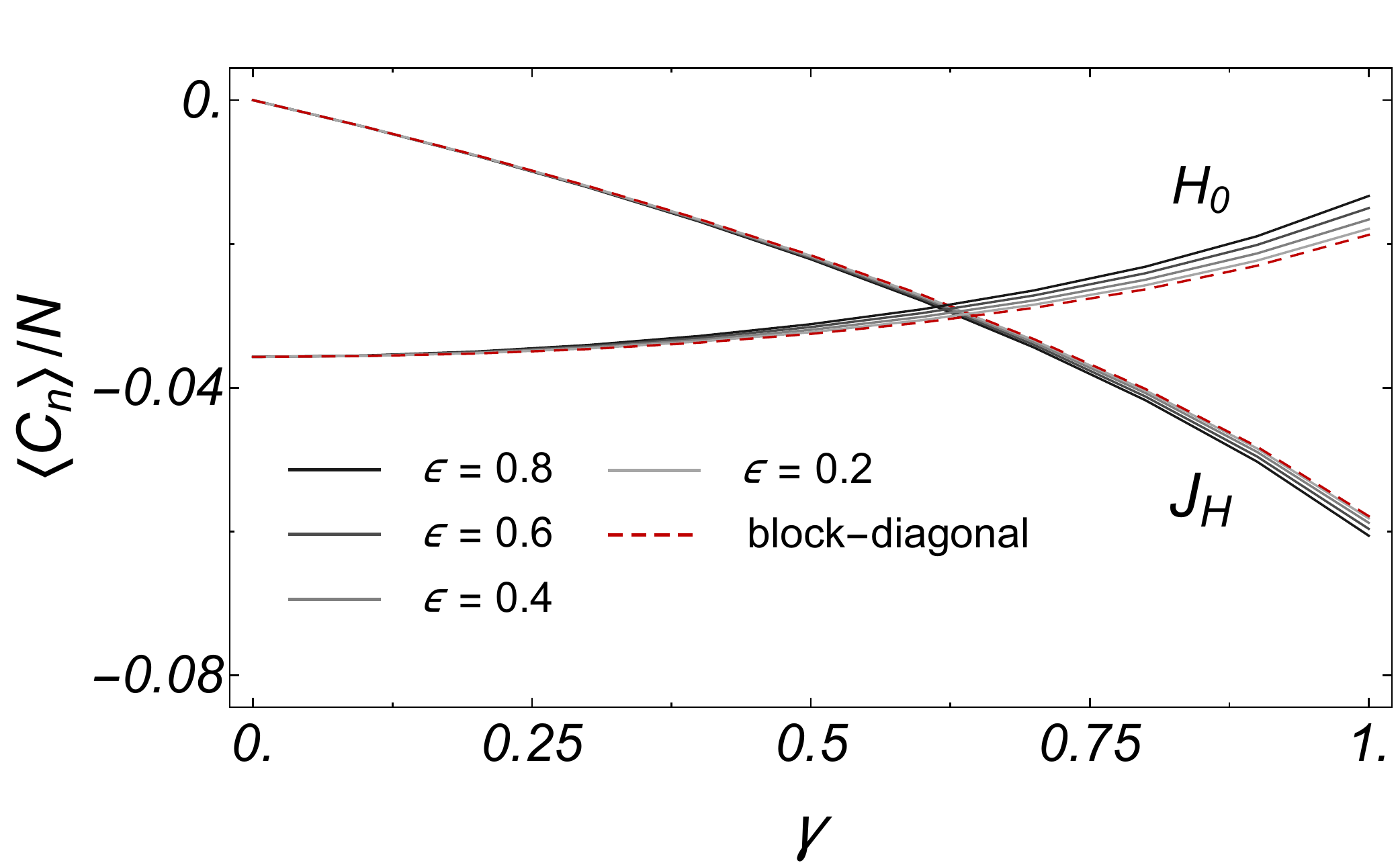}\\
\caption{Steady state expectation values of the energy and the heat current $J_H$ as a function of $\gamma$ parametrizing the type of Markovian coupling. For a system of $N=8$ sites the steady state is shown at $J=1$ for several values of $\epsilon$ and also for the limit $\epsilon\to 0$ where the block-diagonal ensemble becomes exact.
  \label{FigExact}}
\end{figure}

In Fig.~\eqref{FigExact} we show the steady-state expectation value of the energy and the heat current as function of $\gamma$. The figure shows that the block-diagonal density matrix quantitatively describes for all values of $\gamma$ the steady state for moderate values of $\epsilon$.

\newpage

\clearpage


\begin{thebibliography}{63}%
\makeatletter
\providecommand \@ifxundefined [1]{%
 \@ifx{#1\undefined}
}%
\providecommand \@ifnum [1]{%
 \ifnum #1\expandafter \@firstoftwo
 \else \expandafter \@secondoftwo
 \fi
}%
\providecommand \@ifx [1]{%
 \ifx #1\expandafter \@firstoftwo
 \else \expandafter \@secondoftwo
 \fi
}%
\providecommand \natexlab [1]{#1}%
\providecommand \enquote  [1]{``#1''}%
\providecommand \bibnamefont  [1]{#1}%
\providecommand \bibfnamefont [1]{#1}%
\providecommand \citenamefont [1]{#1}%
\providecommand \href@noop [0]{\@secondoftwo}%
\providecommand \href [0]{\begingroup \@sanitize@url \@href}%
\providecommand \@href[1]{\@@startlink{#1}\@@href}%
\providecommand \@@href[1]{\endgroup#1\@@endlink}%
\providecommand \@sanitize@url [0]{\catcode `\\12\catcode `\$12\catcode
  `\&12\catcode `\#12\catcode `\^12\catcode `\_12\catcode `\%12\relax}%
\providecommand \@@startlink[1]{}%
\providecommand \@@endlink[0]{}%
\providecommand \url  [0]{\begingroup\@sanitize@url \@url }%
\providecommand \@url [1]{\endgroup\@href {#1}{\urlprefix }}%
\providecommand \urlprefix  [0]{URL }%
\providecommand \Eprint [0]{\href }%
\providecommand \doibase [0]{http://dx.doi.org/}%
\providecommand \selectlanguage [0]{\@gobble}%
\providecommand \bibinfo  [0]{\@secondoftwo}%
\providecommand \bibfield  [0]{\@secondoftwo}%
\providecommand \translation [1]{[#1]}%
\providecommand \BibitemOpen [0]{}%
\providecommand \bibitemStop [0]{}%
\providecommand \bibitemNoStop [0]{.\EOS\space}%
\providecommand \EOS [0]{\spacefactor3000\relax}%
\providecommand \BibitemShut  [1]{\csname bibitem#1\endcsname}%
\let\auto@bib@innerbib\@empty
\bibitem [{\citenamefont {Polkovnikov}\ \emph {et~al.}(2011)\citenamefont
  {Polkovnikov}, \citenamefont {Sengupta}, \citenamefont {Silva},\ and\
  \citenamefont {Vengalattore}}]{polkovnikov11}%
  \BibitemOpen
  \bibfield  {author} {\bibinfo {author} {\bibfnamefont {A.}~\bibnamefont
  {Polkovnikov}}, \bibinfo {author} {\bibfnamefont {K.}~\bibnamefont
  {Sengupta}}, \bibinfo {author} {\bibfnamefont {A.}~\bibnamefont {Silva}}, \
  and\ \bibinfo {author} {\bibfnamefont {M.}~\bibnamefont {Vengalattore}},\
  }\href {\doibase 10.1103/RevModPhys.83.863} {\bibfield  {journal} {\bibinfo
  {journal} {Rev. Mod. Phys.}\ }\textbf {\bibinfo {volume} {83}},\ \bibinfo
  {pages} {863} (\bibinfo {year} {2011})}\BibitemShut {NoStop}%
\bibitem [{\citenamefont {Gogolin}\ and\ \citenamefont
  {Eisert}(2016)}]{gogolin16}%
  \BibitemOpen
  \bibfield  {author} {\bibinfo {author} {\bibfnamefont {C.}~\bibnamefont
  {Gogolin}}\ and\ \bibinfo {author} {\bibfnamefont {J.}~\bibnamefont
  {Eisert}},\ }\href@noop {} {\bibfield  {journal} {\bibinfo  {journal}
  {Reports on Progress in Physics}\ }\textbf {\bibinfo {volume} {79}},\
  \bibinfo {pages} {056001} (\bibinfo {year} {2016})}\BibitemShut {NoStop}%
\bibitem [{\citenamefont {D'Alessio}\ \emph {et~al.}(2016)\citenamefont
  {D'Alessio}, \citenamefont {Kafri}, \citenamefont {Polkovnikov},\ and\
  \citenamefont {Rigol}}]{dalessio016}%
  \BibitemOpen
  \bibfield  {author} {\bibinfo {author} {\bibfnamefont {L.}~\bibnamefont
  {D'Alessio}}, \bibinfo {author} {\bibfnamefont {Y.}~\bibnamefont {Kafri}},
  \bibinfo {author} {\bibfnamefont {A.}~\bibnamefont {Polkovnikov}}, \ and\
  \bibinfo {author} {\bibfnamefont {M.}~\bibnamefont {Rigol}},\ }\href@noop {}
  {\bibfield  {journal} {\bibinfo  {journal} {Advances in Physics}\ }\textbf
  {\bibinfo {volume} {65}},\ \bibinfo {pages} {239} (\bibinfo {year}
  {2016})}\BibitemShut {NoStop}%
\bibitem [{\citenamefont {Rigol}\ \emph {et~al.}(2008)\citenamefont {Rigol},
  \citenamefont {Dunjko},\ and\ \citenamefont {Olshanii}}]{rigol08}%
  \BibitemOpen
  \bibfield  {author} {\bibinfo {author} {\bibfnamefont {M.}~\bibnamefont
  {Rigol}}, \bibinfo {author} {\bibfnamefont {V.}~\bibnamefont {Dunjko}}, \
  and\ \bibinfo {author} {\bibfnamefont {M.}~\bibnamefont {Olshanii}},\
  }\href@noop {} {\bibfield  {journal} {\bibinfo  {journal} {Nature}\ }\textbf
  {\bibinfo {volume} {452}},\ \bibinfo {pages} {854} (\bibinfo {year}
  {2008})}\BibitemShut {NoStop}%
\bibitem [{\citenamefont {Lux}\ \emph {et~al.}(2014)\citenamefont {Lux},
  \citenamefont {M\"uller}, \citenamefont {Mitra},\ and\ \citenamefont
  {Rosch}}]{lux14}%
  \BibitemOpen
  \bibfield  {author} {\bibinfo {author} {\bibfnamefont {J.}~\bibnamefont
  {Lux}}, \bibinfo {author} {\bibfnamefont {J.}~\bibnamefont {M\"uller}},
  \bibinfo {author} {\bibfnamefont {A.}~\bibnamefont {Mitra}}, \ and\ \bibinfo
  {author} {\bibfnamefont {A.}~\bibnamefont {Rosch}},\ }\href {\doibase
  10.1103/PhysRevA.89.053608} {\bibfield  {journal} {\bibinfo  {journal} {Phys.
  Rev. A}\ }\textbf {\bibinfo {volume} {89}},\ \bibinfo {pages} {053608}
  (\bibinfo {year} {2014})}\BibitemShut {NoStop}%
\bibitem [{\citenamefont {Bohrdt}\ \emph {et~al.}(2017)\citenamefont {Bohrdt},
  \citenamefont {Mendl}, \citenamefont {Endres},\ and\ \citenamefont
  {Knap}}]{bohrdt17}%
  \BibitemOpen
  \bibfield  {author} {\bibinfo {author} {\bibfnamefont {A.}~\bibnamefont
  {Bohrdt}}, \bibinfo {author} {\bibfnamefont {C.}~\bibnamefont {Mendl}},
  \bibinfo {author} {\bibfnamefont {M.}~\bibnamefont {Endres}}, \ and\ \bibinfo
  {author} {\bibfnamefont {M.}~\bibnamefont {Knap}},\ }\href@noop {} {\bibfield
   {journal} {\bibinfo  {journal} {New Journal of Physics}\ } (\bibinfo {year}
  {2017})}\BibitemShut {NoStop}%
\bibitem [{\citenamefont {Leviatan}\ \emph {et~al.}(2017)\citenamefont
  {Leviatan}, \citenamefont {Pollmann}, \citenamefont {Bardarson},\ and\
  \citenamefont {Altman}}]{leviatan17}%
  \BibitemOpen
  \bibfield  {author} {\bibinfo {author} {\bibfnamefont {E.}~\bibnamefont
  {Leviatan}}, \bibinfo {author} {\bibfnamefont {F.}~\bibnamefont {Pollmann}},
  \bibinfo {author} {\bibfnamefont {J.~H.}\ \bibnamefont {Bardarson}}, \ and\
  \bibinfo {author} {\bibfnamefont {E.}~\bibnamefont {Altman}},\ }\href@noop {}
  {\bibfield  {journal} {\bibinfo  {journal} {arXiv:1702.08894}\ } (\bibinfo
  {year} {2017})}\BibitemShut {NoStop}%
\bibitem [{\citenamefont {Essler}\ and\ \citenamefont
  {Fagotti}(2016)}]{essler16}%
  \BibitemOpen
  \bibfield  {author} {\bibinfo {author} {\bibfnamefont {F.~H.~L.}\
  \bibnamefont {Essler}}\ and\ \bibinfo {author} {\bibfnamefont
  {M.}~\bibnamefont {Fagotti}},\ }\href@noop {} {\bibfield  {journal} {\bibinfo
   {journal} {Journal of Statistical Mechanics: Theory and Experiment P064002}\
  } (\bibinfo {year} {2016})}\BibitemShut {NoStop}%
\bibitem [{\citenamefont {Vidmar}\ and\ \citenamefont
  {Rigol}(2016)}]{vidmar16}%
  \BibitemOpen
  \bibfield  {author} {\bibinfo {author} {\bibfnamefont {L.}~\bibnamefont
  {Vidmar}}\ and\ \bibinfo {author} {\bibfnamefont {M.}~\bibnamefont {Rigol}},\
  }\href@noop {} {\bibfield  {journal} {\bibinfo  {journal} {Journal of
  Statistical Mechanics: Theory and Experiment P064007}\ } (\bibinfo {year}
  {2016})}\BibitemShut {NoStop}%
\bibitem [{\citenamefont {Calabrese}\ and\ \citenamefont
  {Cardy}(2016)}]{calabrese16}%
  \BibitemOpen
  \bibfield  {author} {\bibinfo {author} {\bibfnamefont {P.}~\bibnamefont
  {Calabrese}}\ and\ \bibinfo {author} {\bibfnamefont {J.}~\bibnamefont
  {Cardy}},\ }\href {http://stacks.iop.org/1742-5468/2016/i=6/a=064003}
  {\bibfield  {journal} {\bibinfo  {journal} {Journal of Statistical Mechanics:
  Theory and Experiment}\ }\textbf {\bibinfo {volume} {2016}},\ \bibinfo
  {pages} {064003} (\bibinfo {year} {2016})}\BibitemShut {NoStop}%
\bibitem [{\citenamefont {Cazalilla}\ and\ \citenamefont
  {Chung}(2016)}]{cazalilla16}%
  \BibitemOpen
  \bibfield  {author} {\bibinfo {author} {\bibfnamefont {M.~A.}\ \bibnamefont
  {Cazalilla}}\ and\ \bibinfo {author} {\bibfnamefont {M.-C.}\ \bibnamefont
  {Chung}},\ }\href {http://stacks.iop.org/1742-5468/2016/i=6/a=064004}
  {\bibfield  {journal} {\bibinfo  {journal} {Journal of Statistical Mechanics:
  Theory and Experiment}\ }\textbf {\bibinfo {volume} {2016}},\ \bibinfo
  {pages} {064004} (\bibinfo {year} {2016})}\BibitemShut {NoStop}%
\bibitem [{\citenamefont {Rigol}\ \emph {et~al.}(2007)\citenamefont {Rigol},
  \citenamefont {Dunjko}, \citenamefont {Yurovsky},\ and\ \citenamefont
  {Olshanii}}]{rigol07}%
  \BibitemOpen
  \bibfield  {author} {\bibinfo {author} {\bibfnamefont {M.}~\bibnamefont
  {Rigol}}, \bibinfo {author} {\bibfnamefont {V.}~\bibnamefont {Dunjko}},
  \bibinfo {author} {\bibfnamefont {V.}~\bibnamefont {Yurovsky}}, \ and\
  \bibinfo {author} {\bibfnamefont {M.}~\bibnamefont {Olshanii}},\ }\href
  {\doibase 10.1103/PhysRevLett.98.050405} {\bibfield  {journal} {\bibinfo
  {journal} {Phys. Rev. Lett.}\ }\textbf {\bibinfo {volume} {98}},\ \bibinfo
  {pages} {050405} (\bibinfo {year} {2007})}\BibitemShut {NoStop}%
\bibitem [{\citenamefont {Fagotti}\ and\ \citenamefont
  {Essler}(2013{\natexlab{a}})}]{fagotti13gge2}%
  \BibitemOpen
  \bibfield  {author} {\bibinfo {author} {\bibfnamefont {M.}~\bibnamefont
  {Fagotti}}\ and\ \bibinfo {author} {\bibfnamefont {F.~H.~L.}\ \bibnamefont
  {Essler}},\ }\href {\doibase 10.1103/PhysRevB.87.245107} {\bibfield
  {journal} {\bibinfo  {journal} {Phys. Rev. B}\ }\textbf {\bibinfo {volume}
  {87}},\ \bibinfo {pages} {245107} (\bibinfo {year}
  {2013}{\natexlab{a}})}\BibitemShut {NoStop}%
\bibitem [{\citenamefont {Pozsgay}(2013)}]{pozsgay13}%
  \BibitemOpen
  \bibfield  {author} {\bibinfo {author} {\bibfnamefont {B.}~\bibnamefont
  {Pozsgay}},\ }\href@noop {} {\bibfield  {journal} {\bibinfo  {journal}
  {Journal of Statistical Mechanics: Theory and Experiment P07003}\ } (\bibinfo
  {year} {2013})}\BibitemShut {NoStop}%
\bibitem [{\citenamefont {Fagotti}\ and\ \citenamefont
  {Essler}(2013{\natexlab{b}})}]{fagotti13}%
  \BibitemOpen
  \bibfield  {author} {\bibinfo {author} {\bibfnamefont {M.}~\bibnamefont
  {Fagotti}}\ and\ \bibinfo {author} {\bibfnamefont {F.~H.}\ \bibnamefont
  {Essler}},\ }\href@noop {} {\bibfield  {journal} {\bibinfo  {journal}
  {Journal of Statistical Mechanics: Theory and Experiment P07012}\ } (\bibinfo
  {year} {2013}{\natexlab{b}})}\BibitemShut {NoStop}%
\bibitem [{\citenamefont {Ilievski}\ \emph
  {et~al.}(2015{\natexlab{a}})\citenamefont {Ilievski}, \citenamefont
  {De~Nardis}, \citenamefont {Wouters}, \citenamefont {Caux}, \citenamefont
  {Essler},\ and\ \citenamefont {Prosen}}]{ilievski15a}%
  \BibitemOpen
  \bibfield  {author} {\bibinfo {author} {\bibfnamefont {E.}~\bibnamefont
  {Ilievski}}, \bibinfo {author} {\bibfnamefont {J.}~\bibnamefont {De~Nardis}},
  \bibinfo {author} {\bibfnamefont {B.}~\bibnamefont {Wouters}}, \bibinfo
  {author} {\bibfnamefont {J.-S.}\ \bibnamefont {Caux}}, \bibinfo {author}
  {\bibfnamefont {F.~H.~L.}\ \bibnamefont {Essler}}, \ and\ \bibinfo {author}
  {\bibfnamefont {T.}~\bibnamefont {Prosen}},\ }\href {\doibase
  10.1103/PhysRevLett.115.157201} {\bibfield  {journal} {\bibinfo  {journal}
  {Phys. Rev. Lett.}\ }\textbf {\bibinfo {volume} {115}},\ \bibinfo {pages}
  {157201} (\bibinfo {year} {2015}{\natexlab{a}})}\BibitemShut {NoStop}%
\bibitem [{\citenamefont {Ilievski}\ \emph
  {et~al.}(2015{\natexlab{b}})\citenamefont {Ilievski}, \citenamefont
  {Medenjak},\ and\ \citenamefont {Prosen}}]{ilievski15}%
  \BibitemOpen
  \bibfield  {author} {\bibinfo {author} {\bibfnamefont {E.}~\bibnamefont
  {Ilievski}}, \bibinfo {author} {\bibfnamefont {M.}~\bibnamefont {Medenjak}},
  \ and\ \bibinfo {author} {\bibfnamefont {T.}~\bibnamefont {Prosen}},\ }\href
  {\doibase 10.1103/PhysRevLett.115.120601} {\bibfield  {journal} {\bibinfo
  {journal} {Phys. Rev. Lett.}\ }\textbf {\bibinfo {volume} {115}},\ \bibinfo
  {pages} {120601} (\bibinfo {year} {2015}{\natexlab{b}})}\BibitemShut
  {NoStop}%
\bibitem [{\citenamefont {Mierzejewski}\ \emph {et~al.}(2015)\citenamefont
  {Mierzejewski}, \citenamefont {Prelov\v{s}ek},\ and\ \citenamefont
  {Prosen}}]{mierzejewski15}%
  \BibitemOpen
  \bibfield  {author} {\bibinfo {author} {\bibfnamefont {M.}~\bibnamefont
  {Mierzejewski}}, \bibinfo {author} {\bibfnamefont {P.}~\bibnamefont
  {Prelov\v{s}ek}}, \ and\ \bibinfo {author} {\bibfnamefont {T.}~\bibnamefont
  {Prosen}},\ }\href {\doibase 10.1103/PhysRevLett.114.140601} {\bibfield
  {journal} {\bibinfo  {journal} {Phys. Rev. Lett.}\ }\textbf {\bibinfo
  {volume} {114}},\ \bibinfo {pages} {140601} (\bibinfo {year}
  {2015})}\BibitemShut {NoStop}%
\bibitem [{\citenamefont {Ilievski}\ \emph {et~al.}(2016)\citenamefont
  {Ilievski}, \citenamefont {Medenjak}, \citenamefont {Prosen},\ and\
  \citenamefont {Zadnik}}]{ilievski16}%
  \BibitemOpen
  \bibfield  {author} {\bibinfo {author} {\bibfnamefont {E.}~\bibnamefont
  {Ilievski}}, \bibinfo {author} {\bibfnamefont {M.}~\bibnamefont {Medenjak}},
  \bibinfo {author} {\bibfnamefont {T.}~\bibnamefont {Prosen}}, \ and\ \bibinfo
  {author} {\bibfnamefont {L.}~\bibnamefont {Zadnik}},\ }\href@noop {}
  {\bibfield  {journal} {\bibinfo  {journal} {Journal of Statistical Mechanics:
  Theory and Experiment P064008}\ } (\bibinfo {year} {2016})}\BibitemShut
  {NoStop}%
\bibitem [{\citenamefont {Caux}(2016)}]{caux16}%
  \BibitemOpen
  \bibfield  {author} {\bibinfo {author} {\bibfnamefont {J.-S.}\ \bibnamefont
  {Caux}},\ }\href@noop {} {\bibfield  {journal} {\bibinfo  {journal} {Journal
  of Statistical Mechanics: Theory and Experiment P064006}\ } (\bibinfo {year}
  {2016})}\BibitemShut {NoStop}%
\bibitem [{\citenamefont {Caux}\ and\ \citenamefont {Essler}(2013)}]{caux13}%
  \BibitemOpen
  \bibfield  {author} {\bibinfo {author} {\bibfnamefont {J.-S.}\ \bibnamefont
  {Caux}}\ and\ \bibinfo {author} {\bibfnamefont {F.~H.~L.}\ \bibnamefont
  {Essler}},\ }\href {\doibase 10.1103/PhysRevLett.110.257203} {\bibfield
  {journal} {\bibinfo  {journal} {Phys. Rev. Lett.}\ }\textbf {\bibinfo
  {volume} {110}},\ \bibinfo {pages} {257203} (\bibinfo {year}
  {2013})}\BibitemShut {NoStop}%
\bibitem [{\citenamefont {Brockmann}\ \emph {et~al.}(2014)\citenamefont
  {Brockmann}, \citenamefont {Wouters}, \citenamefont {Fioretto}, \citenamefont
  {De~Nardis}, \citenamefont {Vlijm},\ and\ \citenamefont
  {Caux}}]{brockmann14gge}%
  \BibitemOpen
  \bibfield  {author} {\bibinfo {author} {\bibfnamefont {M.}~\bibnamefont
  {Brockmann}}, \bibinfo {author} {\bibfnamefont {B.}~\bibnamefont {Wouters}},
  \bibinfo {author} {\bibfnamefont {D.}~\bibnamefont {Fioretto}}, \bibinfo
  {author} {\bibfnamefont {J.}~\bibnamefont {De~Nardis}}, \bibinfo {author}
  {\bibfnamefont {R.}~\bibnamefont {Vlijm}}, \ and\ \bibinfo {author}
  {\bibfnamefont {J.-S.}\ \bibnamefont {Caux}},\ }\href
  {http://stacks.iop.org/1742-5468/2014/i=12/a=P12009} {\bibfield  {journal}
  {\bibinfo  {journal} {Journal of Statistical Mechanics: Theory and
  Experiment}\ }\textbf {\bibinfo {volume} {2014}},\ \bibinfo {pages} {P12009}
  (\bibinfo {year} {2014})}\BibitemShut {NoStop}%
\bibitem [{\citenamefont {Wouters}\ \emph {et~al.}(2014)\citenamefont
  {Wouters}, \citenamefont {De~Nardis}, \citenamefont {Brockmann},
  \citenamefont {Fioretto}, \citenamefont {Rigol},\ and\ \citenamefont
  {Caux}}]{wouters14}%
  \BibitemOpen
  \bibfield  {author} {\bibinfo {author} {\bibfnamefont {B.}~\bibnamefont
  {Wouters}}, \bibinfo {author} {\bibfnamefont {J.}~\bibnamefont {De~Nardis}},
  \bibinfo {author} {\bibfnamefont {M.}~\bibnamefont {Brockmann}}, \bibinfo
  {author} {\bibfnamefont {D.}~\bibnamefont {Fioretto}}, \bibinfo {author}
  {\bibfnamefont {M.}~\bibnamefont {Rigol}}, \ and\ \bibinfo {author}
  {\bibfnamefont {J.-S.}\ \bibnamefont {Caux}},\ }\href {\doibase
  10.1103/PhysRevLett.113.117202} {\bibfield  {journal} {\bibinfo  {journal}
  {Phys. Rev. Lett.}\ }\textbf {\bibinfo {volume} {113}},\ \bibinfo {pages}
  {117202} (\bibinfo {year} {2014})}\BibitemShut {NoStop}%
\bibitem [{\citenamefont {Pozsgay}\ \emph {et~al.}(2014)\citenamefont
  {Pozsgay}, \citenamefont {Mesty\'an}, \citenamefont {Werner}, \citenamefont
  {Kormos}, \citenamefont {Zar\'and},\ and\ \citenamefont
  {Tak\'acs}}]{pozsgay14}%
  \BibitemOpen
  \bibfield  {author} {\bibinfo {author} {\bibfnamefont {B.}~\bibnamefont
  {Pozsgay}}, \bibinfo {author} {\bibfnamefont {M.}~\bibnamefont {Mesty\'an}},
  \bibinfo {author} {\bibfnamefont {M.~A.}\ \bibnamefont {Werner}}, \bibinfo
  {author} {\bibfnamefont {M.}~\bibnamefont {Kormos}}, \bibinfo {author}
  {\bibfnamefont {G.}~\bibnamefont {Zar\'and}}, \ and\ \bibinfo {author}
  {\bibfnamefont {G.}~\bibnamefont {Tak\'acs}},\ }\href {\doibase
  10.1103/PhysRevLett.113.117203} {\bibfield  {journal} {\bibinfo  {journal}
  {Phys. Rev. Lett.}\ }\textbf {\bibinfo {volume} {113}},\ \bibinfo {pages}
  {117203} (\bibinfo {year} {2014})}\BibitemShut {NoStop}%
\bibitem [{\citenamefont {Mestyán}\ \emph {et~al.}(2015)\citenamefont
  {Mestyán}, \citenamefont {Pozsgay}, \citenamefont {Tak\'acs},\ and\
  \citenamefont {Werner}}]{mestyan15gge}%
  \BibitemOpen
  \bibfield  {author} {\bibinfo {author} {\bibfnamefont {M.}~\bibnamefont
  {Mestyán}}, \bibinfo {author} {\bibfnamefont {B.}~\bibnamefont {Pozsgay}},
  \bibinfo {author} {\bibfnamefont {G.}~\bibnamefont {Tak\'acs}}, \ and\
  \bibinfo {author} {\bibfnamefont {M.~A.}\ \bibnamefont {Werner}},\ }\href
  {http://stacks.iop.org/1742-5468/2015/i=4/a=P04001} {\bibfield  {journal}
  {\bibinfo  {journal} {Journal of Statistical Mechanics: Theory and
  Experiment}\ }\textbf {\bibinfo {volume} {2015}},\ \bibinfo {pages} {P04001}
  (\bibinfo {year} {2015})}\BibitemShut {NoStop}%
\bibitem [{\citenamefont {Piroli}\ \emph {et~al.}(2016)\citenamefont {Piroli},
  \citenamefont {Vernier},\ and\ \citenamefont {Calabrese}}]{piroli16gge}%
  \BibitemOpen
  \bibfield  {author} {\bibinfo {author} {\bibfnamefont {L.}~\bibnamefont
  {Piroli}}, \bibinfo {author} {\bibfnamefont {E.}~\bibnamefont {Vernier}}, \
  and\ \bibinfo {author} {\bibfnamefont {P.}~\bibnamefont {Calabrese}},\ }\href
  {\doibase 10.1103/PhysRevB.94.054313} {\bibfield  {journal} {\bibinfo
  {journal} {Phys. Rev. B}\ }\textbf {\bibinfo {volume} {94}},\ \bibinfo
  {pages} {054313} (\bibinfo {year} {2016})}\BibitemShut {NoStop}%
\bibitem [{\citenamefont {Pozsgay}\ \emph {et~al.}(2017)\citenamefont
  {Pozsgay}, \citenamefont {Vernier},\ and\ \citenamefont
  {Werner}}]{pozsgay17}%
  \BibitemOpen
  \bibfield  {author} {\bibinfo {author} {\bibfnamefont {B.}~\bibnamefont
  {Pozsgay}}, \bibinfo {author} {\bibfnamefont {E.}~\bibnamefont {Vernier}}, \
  and\ \bibinfo {author} {\bibfnamefont {M.}~\bibnamefont {Werner}},\
  }\href@noop {} {\bibfield  {journal} {\bibinfo  {journal} {arXiv preprint
  arXiv:1703.09516}\ } (\bibinfo {year} {2017})}\BibitemShut {NoStop}%
\bibitem [{\citenamefont {Kinoshita}\ \emph {et~al.}(2006)\citenamefont
  {Kinoshita}, \citenamefont {Wenger},\ and\ \citenamefont
  {Weiss}}]{kinoshita06}%
  \BibitemOpen
  \bibfield  {author} {\bibinfo {author} {\bibfnamefont {T.}~\bibnamefont
  {Kinoshita}}, \bibinfo {author} {\bibfnamefont {T.}~\bibnamefont {Wenger}}, \
  and\ \bibinfo {author} {\bibfnamefont {D.~S.}\ \bibnamefont {Weiss}},\
  }\href@noop {} {\bibfield  {journal} {\bibinfo  {journal} {Nature}\ }\textbf
  {\bibinfo {volume} {440}},\ \bibinfo {pages} {900} (\bibinfo {year}
  {2006})}\BibitemShut {NoStop}%
\bibitem [{\citenamefont {Gring}\ \emph {et~al.}(2012)\citenamefont {Gring},
  \citenamefont {Kuhnert}, \citenamefont {Langen}, \citenamefont {Kitagawa},
  \citenamefont {Rauer}, \citenamefont {Schreitl}, \citenamefont {Mazets},
  \citenamefont {Smith}, \citenamefont {Demler},\ and\ \citenamefont
  {Schmiedmayer}}]{gring12}%
  \BibitemOpen
  \bibfield  {author} {\bibinfo {author} {\bibfnamefont {M.}~\bibnamefont
  {Gring}}, \bibinfo {author} {\bibfnamefont {M.}~\bibnamefont {Kuhnert}},
  \bibinfo {author} {\bibfnamefont {T.}~\bibnamefont {Langen}}, \bibinfo
  {author} {\bibfnamefont {T.}~\bibnamefont {Kitagawa}}, \bibinfo {author}
  {\bibfnamefont {B.}~\bibnamefont {Rauer}}, \bibinfo {author} {\bibfnamefont
  {M.}~\bibnamefont {Schreitl}}, \bibinfo {author} {\bibfnamefont
  {I.}~\bibnamefont {Mazets}}, \bibinfo {author} {\bibfnamefont {D.~A.}\
  \bibnamefont {Smith}}, \bibinfo {author} {\bibfnamefont {E.}~\bibnamefont
  {Demler}}, \ and\ \bibinfo {author} {\bibfnamefont {J.}~\bibnamefont
  {Schmiedmayer}},\ }\href@noop {} {\bibfield  {journal} {\bibinfo  {journal}
  {Science}\ }\textbf {\bibinfo {volume} {337}},\ \bibinfo {pages} {1318}
  (\bibinfo {year} {2012})}\BibitemShut {NoStop}%
\bibitem [{\citenamefont {Smith}\ \emph {et~al.}(2013)\citenamefont {Smith},
  \citenamefont {Gring}, \citenamefont {Langen}, \citenamefont {Kuhnert},
  \citenamefont {Rauer}, \citenamefont {Geiger}, \citenamefont {Kitagawa},
  \citenamefont {Mazets}, \citenamefont {Demler},\ and\ \citenamefont
  {Schmiedmayer}}]{smith13}%
  \BibitemOpen
  \bibfield  {author} {\bibinfo {author} {\bibfnamefont {D.~A.}\ \bibnamefont
  {Smith}}, \bibinfo {author} {\bibfnamefont {M.}~\bibnamefont {Gring}},
  \bibinfo {author} {\bibfnamefont {T.}~\bibnamefont {Langen}}, \bibinfo
  {author} {\bibfnamefont {M.}~\bibnamefont {Kuhnert}}, \bibinfo {author}
  {\bibfnamefont {B.}~\bibnamefont {Rauer}}, \bibinfo {author} {\bibfnamefont
  {R.}~\bibnamefont {Geiger}}, \bibinfo {author} {\bibfnamefont
  {T.}~\bibnamefont {Kitagawa}}, \bibinfo {author} {\bibfnamefont
  {I.}~\bibnamefont {Mazets}}, \bibinfo {author} {\bibfnamefont
  {E.}~\bibnamefont {Demler}}, \ and\ \bibinfo {author} {\bibfnamefont
  {J.}~\bibnamefont {Schmiedmayer}},\ }\href@noop {} {\bibfield  {journal}
  {\bibinfo  {journal} {New Journal of Physics}\ }\textbf {\bibinfo {volume}
  {15}},\ \bibinfo {pages} {075011} (\bibinfo {year} {2013})}\BibitemShut
  {NoStop}%
\bibitem [{\citenamefont {Langen}\ \emph {et~al.}(2016)\citenamefont {Langen},
  \citenamefont {Gasenzer},\ and\ \citenamefont {Schmiedmayer}}]{Langen2016}%
  \BibitemOpen
  \bibfield  {author} {\bibinfo {author} {\bibfnamefont {T.}~\bibnamefont
  {Langen}}, \bibinfo {author} {\bibfnamefont {T.}~\bibnamefont {Gasenzer}}, \
  and\ \bibinfo {author} {\bibfnamefont {J.}~\bibnamefont {Schmiedmayer}},\
  }\href {http://stacks.iop.org/1742-5468/2016/i=6/a=064009} {\bibfield
  {journal} {\bibinfo  {journal} {J. Stat. Mech.}\ }\textbf {\bibinfo {volume}
  {2016}},\ \bibinfo {pages} {064009} (\bibinfo {year} {2016})}\BibitemShut
  {NoStop}%
\bibitem [{\citenamefont {Langen}\ \emph {et~al.}(2015)\citenamefont {Langen},
  \citenamefont {Erne}, \citenamefont {Geiger}, \citenamefont {Rauer},
  \citenamefont {Schweigler}, \citenamefont {Kuhnert}, \citenamefont
  {Rohringer}, \citenamefont {Mazets}, \citenamefont {Gasenzer},\ and\
  \citenamefont {Schmiedmayer}}]{langen15}%
  \BibitemOpen
  \bibfield  {author} {\bibinfo {author} {\bibfnamefont {T.}~\bibnamefont
  {Langen}}, \bibinfo {author} {\bibfnamefont {S.}~\bibnamefont {Erne}},
  \bibinfo {author} {\bibfnamefont {R.}~\bibnamefont {Geiger}}, \bibinfo
  {author} {\bibfnamefont {B.}~\bibnamefont {Rauer}}, \bibinfo {author}
  {\bibfnamefont {T.}~\bibnamefont {Schweigler}}, \bibinfo {author}
  {\bibfnamefont {M.}~\bibnamefont {Kuhnert}}, \bibinfo {author} {\bibfnamefont
  {W.}~\bibnamefont {Rohringer}}, \bibinfo {author} {\bibfnamefont {I.~E.}\
  \bibnamefont {Mazets}}, \bibinfo {author} {\bibfnamefont {T.}~\bibnamefont
  {Gasenzer}}, \ and\ \bibinfo {author} {\bibfnamefont {J.}~\bibnamefont
  {Schmiedmayer}},\ }\href@noop {} {\bibfield  {journal} {\bibinfo  {journal}
  {Science}\ }\textbf {\bibinfo {volume} {348}},\ \bibinfo {pages} {207}
  (\bibinfo {year} {2015})}\BibitemShut {NoStop}%
\bibitem [{\citenamefont {Tang}\ \emph {et~al.}(2017)\citenamefont {Tang},
  \citenamefont {Kao}, \citenamefont {Li}, \citenamefont {Seo}, \citenamefont
  {Mallayya}, \citenamefont {Rigol}, \citenamefont {Gopalakrishnan},\ and\
  \citenamefont {Lev}}]{tang17}%
  \BibitemOpen
  \bibfield  {author} {\bibinfo {author} {\bibfnamefont {Y.}~\bibnamefont
  {Tang}}, \bibinfo {author} {\bibfnamefont {W.}~\bibnamefont {Kao}}, \bibinfo
  {author} {\bibfnamefont {K.-Y.}\ \bibnamefont {Li}}, \bibinfo {author}
  {\bibfnamefont {S.}~\bibnamefont {Seo}}, \bibinfo {author} {\bibfnamefont
  {K.}~\bibnamefont {Mallayya}}, \bibinfo {author} {\bibfnamefont
  {M.}~\bibnamefont {Rigol}}, \bibinfo {author} {\bibfnamefont
  {S.}~\bibnamefont {Gopalakrishnan}}, \ and\ \bibinfo {author} {\bibfnamefont
  {B.~L.}\ \bibnamefont {Lev}},\ }\href@noop {} {\bibfield  {journal} {\bibinfo
   {journal} {arXiv preprint arXiv:1707.07031}\ } (\bibinfo {year}
  {2017})}\BibitemShut {NoStop}%
\bibitem [{\citenamefont {Jung}\ \emph {et~al.}(2006)\citenamefont {Jung},
  \citenamefont {Helmes},\ and\ \citenamefont {Rosch}}]{jung06}%
  \BibitemOpen
  \bibfield  {author} {\bibinfo {author} {\bibfnamefont {P.}~\bibnamefont
  {Jung}}, \bibinfo {author} {\bibfnamefont {R.~W.}\ \bibnamefont {Helmes}}, \
  and\ \bibinfo {author} {\bibfnamefont {A.}~\bibnamefont {Rosch}},\ }\href
  {\doibase 10.1103/PhysRevLett.96.067202} {\bibfield  {journal} {\bibinfo
  {journal} {Phys. Rev. Lett.}\ }\textbf {\bibinfo {volume} {96}},\ \bibinfo
  {pages} {067202} (\bibinfo {year} {2006})}\BibitemShut {NoStop}%
\bibitem [{\citenamefont {Jung}\ and\ \citenamefont {Rosch}(2007)}]{jung07}%
  \BibitemOpen
  \bibfield  {author} {\bibinfo {author} {\bibfnamefont {P.}~\bibnamefont
  {Jung}}\ and\ \bibinfo {author} {\bibfnamefont {A.}~\bibnamefont {Rosch}},\
  }\href {\doibase 10.1103/PhysRevB.76.245108} {\bibfield  {journal} {\bibinfo
  {journal} {Phys. Rev. B}\ }\textbf {\bibinfo {volume} {76}},\ \bibinfo
  {pages} {245108} (\bibinfo {year} {2007})}\BibitemShut {NoStop}%
\bibitem [{\citenamefont {Bertini}\ \emph {et~al.}(2015)\citenamefont
  {Bertini}, \citenamefont {Essler}, \citenamefont {Groha},\ and\ \citenamefont
  {Robinson}}]{bertini15}%
  \BibitemOpen
  \bibfield  {author} {\bibinfo {author} {\bibfnamefont {B.}~\bibnamefont
  {Bertini}}, \bibinfo {author} {\bibfnamefont {F.~H.~L.}\ \bibnamefont
  {Essler}}, \bibinfo {author} {\bibfnamefont {S.}~\bibnamefont {Groha}}, \
  and\ \bibinfo {author} {\bibfnamefont {N.~J.}\ \bibnamefont {Robinson}},\
  }\href {\doibase 10.1103/PhysRevLett.115.180601} {\bibfield  {journal}
  {\bibinfo  {journal} {Phys. Rev. Lett.}\ }\textbf {\bibinfo {volume} {115}},\
  \bibinfo {pages} {180601} (\bibinfo {year} {2015})}\BibitemShut {NoStop}%
\bibitem [{\citenamefont {Eckstein}\ \emph {et~al.}(2009)\citenamefont
  {Eckstein}, \citenamefont {Kollar},\ and\ \citenamefont
  {Werner}}]{eckstein09}%
  \BibitemOpen
  \bibfield  {author} {\bibinfo {author} {\bibfnamefont {M.}~\bibnamefont
  {Eckstein}}, \bibinfo {author} {\bibfnamefont {M.}~\bibnamefont {Kollar}}, \
  and\ \bibinfo {author} {\bibfnamefont {P.}~\bibnamefont {Werner}},\ }\href
  {\doibase 10.1103/PhysRevLett.103.056403} {\bibfield  {journal} {\bibinfo
  {journal} {Phys. Rev. Lett.}\ }\textbf {\bibinfo {volume} {103}},\ \bibinfo
  {pages} {056403} (\bibinfo {year} {2009})}\BibitemShut {NoStop}%
\bibitem [{\citenamefont {Stark}\ and\ \citenamefont {Kollar}(2013)}]{stark13}%
  \BibitemOpen
  \bibfield  {author} {\bibinfo {author} {\bibfnamefont {M.}~\bibnamefont
  {Stark}}\ and\ \bibinfo {author} {\bibfnamefont {M.}~\bibnamefont {Kollar}},\
  }\href@noop {} {\bibfield  {journal} {\bibinfo  {journal} {arXiv:1308.1610}\
  } (\bibinfo {year} {2013})}\BibitemShut {NoStop}%
\bibitem [{\citenamefont {Marino}\ and\ \citenamefont
  {Silva}(2012)}]{Marino2012}%
  \BibitemOpen
  \bibfield  {author} {\bibinfo {author} {\bibfnamefont {J.}~\bibnamefont
  {Marino}}\ and\ \bibinfo {author} {\bibfnamefont {A.}~\bibnamefont {Silva}},\
  }\href {\doibase 10.1103/PhysRevB.86.060408} {\bibfield  {journal} {\bibinfo
  {journal} {Phys. Rev. B}\ }\textbf {\bibinfo {volume} {86}},\ \bibinfo
  {pages} {060408} (\bibinfo {year} {2012})}\BibitemShut {NoStop}%
\bibitem [{\citenamefont {Marcuzzi}\ \emph {et~al.}(2013)\citenamefont
  {Marcuzzi}, \citenamefont {Marino}, \citenamefont {Gambassi},\ and\
  \citenamefont {Silva}}]{Marcuzzi2013}%
  \BibitemOpen
  \bibfield  {author} {\bibinfo {author} {\bibfnamefont {M.}~\bibnamefont
  {Marcuzzi}}, \bibinfo {author} {\bibfnamefont {J.}~\bibnamefont {Marino}},
  \bibinfo {author} {\bibfnamefont {A.}~\bibnamefont {Gambassi}}, \ and\
  \bibinfo {author} {\bibfnamefont {A.}~\bibnamefont {Silva}},\ }\href
  {\doibase 10.1103/PhysRevLett.111.197203} {\bibfield  {journal} {\bibinfo
  {journal} {Phys. Rev. Lett.}\ }\textbf {\bibinfo {volume} {111}},\ \bibinfo
  {pages} {197203} (\bibinfo {year} {2013})}\BibitemShut {NoStop}%
\bibitem [{\citenamefont {Bertini}\ \emph
  {et~al.}(2016{\natexlab{a}})\citenamefont {Bertini}, \citenamefont {Essler},
  \citenamefont {Groha},\ and\ \citenamefont {Robinson}}]{bertini16}%
  \BibitemOpen
  \bibfield  {author} {\bibinfo {author} {\bibfnamefont {B.}~\bibnamefont
  {Bertini}}, \bibinfo {author} {\bibfnamefont {F.~H.~L.}\ \bibnamefont
  {Essler}}, \bibinfo {author} {\bibfnamefont {S.}~\bibnamefont {Groha}}, \
  and\ \bibinfo {author} {\bibfnamefont {N.~J.}\ \bibnamefont {Robinson}},\
  }\href {\doibase 10.1103/PhysRevB.94.245117} {\bibfield  {journal} {\bibinfo
  {journal} {Phys. Rev. B}\ }\textbf {\bibinfo {volume} {94}},\ \bibinfo
  {pages} {245117} (\bibinfo {year} {2016}{\natexlab{a}})}\BibitemShut
  {NoStop}%
\bibitem [{\citenamefont {Buchhold}\ \emph {et~al.}(2016)\citenamefont
  {Buchhold}, \citenamefont {Heyl},\ and\ \citenamefont {Diehl}}]{buchhold16}%
  \BibitemOpen
  \bibfield  {author} {\bibinfo {author} {\bibfnamefont {M.}~\bibnamefont
  {Buchhold}}, \bibinfo {author} {\bibfnamefont {M.}~\bibnamefont {Heyl}}, \
  and\ \bibinfo {author} {\bibfnamefont {S.}~\bibnamefont {Diehl}},\ }\href
  {\doibase 10.1103/PhysRevA.94.013601} {\bibfield  {journal} {\bibinfo
  {journal} {Phys. Rev. A}\ }\textbf {\bibinfo {volume} {94}},\ \bibinfo
  {pages} {013601} (\bibinfo {year} {2016})}\BibitemShut {NoStop}%
\bibitem [{\citenamefont {Biebl}\ and\ \citenamefont
  {Kehrein}(2017)}]{biebl17}%
  \BibitemOpen
  \bibfield  {author} {\bibinfo {author} {\bibfnamefont {F.~R.~A.}\
  \bibnamefont {Biebl}}\ and\ \bibinfo {author} {\bibfnamefont
  {S.}~\bibnamefont {Kehrein}},\ }\href {\doibase 10.1103/PhysRevB.95.104304}
  {\bibfield  {journal} {\bibinfo  {journal} {Phys. Rev. B}\ }\textbf {\bibinfo
  {volume} {95}},\ \bibinfo {pages} {104304} (\bibinfo {year}
  {2017})}\BibitemShut {NoStop}%
\bibitem [{\citenamefont {Moeckel}\ and\ \citenamefont
  {Kehrein}(2008)}]{moeckel08}%
  \BibitemOpen
  \bibfield  {author} {\bibinfo {author} {\bibfnamefont {M.}~\bibnamefont
  {Moeckel}}\ and\ \bibinfo {author} {\bibfnamefont {S.}~\bibnamefont
  {Kehrein}},\ }\href {\doibase 10.1103/PhysRevLett.100.175702} {\bibfield
  {journal} {\bibinfo  {journal} {Phys. Rev. Lett.}\ }\textbf {\bibinfo
  {volume} {100}},\ \bibinfo {pages} {175702} (\bibinfo {year}
  {2008})}\BibitemShut {NoStop}%
\bibitem [{\citenamefont {Kollar}\ \emph {et~al.}(2011)\citenamefont {Kollar},
  \citenamefont {Wolf},\ and\ \citenamefont {Eckstein}}]{kollar11}%
  \BibitemOpen
  \bibfield  {author} {\bibinfo {author} {\bibfnamefont {M.}~\bibnamefont
  {Kollar}}, \bibinfo {author} {\bibfnamefont {F.~A.}\ \bibnamefont {Wolf}}, \
  and\ \bibinfo {author} {\bibfnamefont {M.}~\bibnamefont {Eckstein}},\
  }\href@noop {} {\bibfield  {journal} {\bibinfo  {journal} {Physical Review
  B}\ }\textbf {\bibinfo {volume} {84}},\ \bibinfo {pages} {054304} (\bibinfo
  {year} {2011})}\BibitemShut {NoStop}%
\bibitem [{\citenamefont {Essler}\ \emph {et~al.}(2014)\citenamefont {Essler},
  \citenamefont {Kehrein}, \citenamefont {Manmana},\ and\ \citenamefont
  {Robinson}}]{essler14}%
  \BibitemOpen
  \bibfield  {author} {\bibinfo {author} {\bibfnamefont {F.~H.~L.}\
  \bibnamefont {Essler}}, \bibinfo {author} {\bibfnamefont {S.}~\bibnamefont
  {Kehrein}}, \bibinfo {author} {\bibfnamefont {S.~R.}\ \bibnamefont
  {Manmana}}, \ and\ \bibinfo {author} {\bibfnamefont {N.~J.}\ \bibnamefont
  {Robinson}},\ }\href {\doibase 10.1103/PhysRevB.89.165104} {\bibfield
  {journal} {\bibinfo  {journal} {Phys. Rev. B}\ }\textbf {\bibinfo {volume}
  {89}},\ \bibinfo {pages} {165104} (\bibinfo {year} {2014})}\BibitemShut
  {NoStop}%
\bibitem [{\citenamefont {Lange}\ \emph {et~al.}(2017)\citenamefont {Lange},
  \citenamefont {Lenar{\v{c}}i{\v{c}}},\ and\ \citenamefont {Rosch}}]{lange17}%
  \BibitemOpen
  \bibfield  {author} {\bibinfo {author} {\bibfnamefont {F.}~\bibnamefont
  {Lange}}, \bibinfo {author} {\bibfnamefont {Z.}~\bibnamefont
  {Lenar{\v{c}}i{\v{c}}}}, \ and\ \bibinfo {author} {\bibfnamefont
  {A.}~\bibnamefont {Rosch}},\ }\href@noop {} {\bibfield  {journal} {\bibinfo
  {journal} {Nature communications}\ }\textbf {\bibinfo {volume} {8}},\
  \bibinfo {pages} {15767} (\bibinfo {year} {2017})}\BibitemShut {NoStop}%
\bibitem [{\citenamefont {Fagotti}(2014)}]{fagotti14tdep}%
  \BibitemOpen
  \bibfield  {author} {\bibinfo {author} {\bibfnamefont {M.}~\bibnamefont
  {Fagotti}},\ }\href {http://stacks.iop.org/1742-5468/2014/i=3/a=P03016}
  {\bibfield  {journal} {\bibinfo  {journal} {Journal of Statistical Mechanics:
  Theory and Experiment}\ }\textbf {\bibinfo {volume} {2014}},\ \bibinfo
  {pages} {P03016} (\bibinfo {year} {2014})}\BibitemShut {NoStop}%
\bibitem [{\citenamefont {Bertini}\ and\ \citenamefont
  {Fagotti}(2015)}]{bertini15a}%
  \BibitemOpen
  \bibfield  {author} {\bibinfo {author} {\bibfnamefont {B.}~\bibnamefont
  {Bertini}}\ and\ \bibinfo {author} {\bibfnamefont {M.}~\bibnamefont
  {Fagotti}},\ }\href {http://stacks.iop.org/1742-5468/2015/i=7/a=P07012}
  {\bibfield  {journal} {\bibinfo  {journal} {Journal of Statistical Mechanics:
  Theory and Experiment}\ }\textbf {\bibinfo {volume} {2015}},\ \bibinfo
  {pages} {P07012} (\bibinfo {year} {2015})}\BibitemShut {NoStop}%
\bibitem [{\citenamefont {Genske}\ and\ \citenamefont
  {Rosch}(2015)}]{genske15}%
  \BibitemOpen
  \bibfield  {author} {\bibinfo {author} {\bibfnamefont {M.}~\bibnamefont
  {Genske}}\ and\ \bibinfo {author} {\bibfnamefont {A.}~\bibnamefont {Rosch}},\
  }\href {\doibase 10.1103/PhysRevA.92.062108} {\bibfield  {journal} {\bibinfo
  {journal} {Phys. Rev. A}\ }\textbf {\bibinfo {volume} {92}},\ \bibinfo
  {pages} {062108} (\bibinfo {year} {2015})}\BibitemShut {NoStop}%
\bibitem [{\citenamefont {Lenar\v{c}i\v{c}}\ \emph {et~al.}(2018)\citenamefont
  {Lenar\v{c}i\v{c}}, \citenamefont {Lange},\ and\ \citenamefont
  {Rosch}}]{lenarcic18}%
  \BibitemOpen
  \bibfield  {author} {\bibinfo {author} {\bibfnamefont {Z.}~\bibnamefont
  {Lenar\v{c}i\v{c}}}, \bibinfo {author} {\bibfnamefont {F.}~\bibnamefont
  {Lange}}, \ and\ \bibinfo {author} {\bibfnamefont {A.}~\bibnamefont
  {Rosch}},\ }\href {\doibase 10.1103/PhysRevB.97.024302} {\bibfield  {journal}
  {\bibinfo  {journal} {Phys. Rev. B}\ }\textbf {\bibinfo {volume} {97}},\
  \bibinfo {pages} {024302} (\bibinfo {year} {2018})}\BibitemShut {NoStop}%
\bibitem [{\citenamefont {Grabowski}\ and\ \citenamefont
  {Mathieu}(1996)}]{grabowski96}%
  \BibitemOpen
  \bibfield  {author} {\bibinfo {author} {\bibfnamefont {M.~P.}\ \bibnamefont
  {Grabowski}}\ and\ \bibinfo {author} {\bibfnamefont {P.}~\bibnamefont
  {Mathieu}},\ }\href {http://stacks.iop.org/0305-4470/29/i=23/a=024}
  {\bibfield  {journal} {\bibinfo  {journal} {Journal of Physics A:
  Mathematical and General}\ }\textbf {\bibinfo {volume} {29}},\ \bibinfo
  {pages} {7635} (\bibinfo {year} {1996})}\BibitemShut {NoStop}%
\bibitem [{\citenamefont {Robertson}(1966)}]{Robertson66}%
  \BibitemOpen
  \bibfield  {author} {\bibinfo {author} {\bibfnamefont {B.}~\bibnamefont
  {Robertson}},\ }\href {\doibase 10.1103/PhysRev.144.151} {\bibfield
  {journal} {\bibinfo  {journal} {Phys. Rev.}\ }\textbf {\bibinfo {volume}
  {144}},\ \bibinfo {pages} {151} (\bibinfo {year} {1966})}\BibitemShut
  {NoStop}%
\bibitem [{\citenamefont {Robertson}(1968)}]{Robertson68}%
  \BibitemOpen
  \bibfield  {author} {\bibinfo {author} {\bibfnamefont {B.}~\bibnamefont
  {Robertson}},\ }\href {\doibase 10.1103/PhysRev.160.175} {\bibfield
  {journal} {\bibinfo  {journal} {Phys. Rev.}\ }\textbf {\bibinfo {volume}
  {160}},\ \bibinfo {pages} {206} (\bibinfo {year} {1968})}\BibitemShut
  {NoStop}%
\bibitem [{\citenamefont {Kawasaki}\ and\ \citenamefont
  {Gunton}(1973)}]{kawasaki73}%
  \BibitemOpen
  \bibfield  {author} {\bibinfo {author} {\bibfnamefont {K.}~\bibnamefont
  {Kawasaki}}\ and\ \bibinfo {author} {\bibfnamefont {J.~D.}\ \bibnamefont
  {Gunton}},\ }\href {\doibase 10.1103/PhysRevA.8.2048} {\bibfield  {journal}
  {\bibinfo  {journal} {Phys. Rev. A}\ }\textbf {\bibinfo {volume} {8}},\
  \bibinfo {pages} {2048} (\bibinfo {year} {1973})}\BibitemShut {NoStop}%
\bibitem [{\citenamefont {Koide}(2002)}]{Koide02}%
  \BibitemOpen
  \bibfield  {author} {\bibinfo {author} {\bibfnamefont {T.}~\bibnamefont
  {Koide}},\ }\href {\doibase 10.1143/PTP.107.525} {\bibfield  {journal}
  {\bibinfo  {journal} {Prog. Theor. Phys.}\ }\textbf {\bibinfo {volume}
  {107}},\ \bibinfo {pages} {525} (\bibinfo {year} {2002})}\BibitemShut
  {NoStop}%
\bibitem [{\citenamefont {Breuer}\ and\ \citenamefont
  {Petruccione}(2002)}]{breuer02}%
  \BibitemOpen
  \bibfield  {author} {\bibinfo {author} {\bibfnamefont {H.~P.}\ \bibnamefont
  {Breuer}}\ and\ \bibinfo {author} {\bibfnamefont {F.}~\bibnamefont
  {Petruccione}},\ }\href@noop {} {\emph {\bibinfo {title} {The theory of open
  quantum systems}}}\ (\bibinfo  {publisher} {Oxford University Press},\
  \bibinfo {address} {New York},\ \bibinfo {year} {2002})\BibitemShut {NoStop}%
\bibitem [{\citenamefont {Barreiro}\ \emph {et~al.}(2011)\citenamefont
  {Barreiro}, \citenamefont {M{\"u}ller}, \citenamefont {Schindler},
  \citenamefont {Nigg}, \citenamefont {Monz}, \citenamefont {Chwalla},
  \citenamefont {Hennrich}, \citenamefont {Roos}, \citenamefont {Zoller},\ and\
  \citenamefont {Blatt}}]{barreiro11}%
  \BibitemOpen
  \bibfield  {author} {\bibinfo {author} {\bibfnamefont {J.~T.}\ \bibnamefont
  {Barreiro}}, \bibinfo {author} {\bibfnamefont {M.}~\bibnamefont
  {M{\"u}ller}}, \bibinfo {author} {\bibfnamefont {P.}~\bibnamefont
  {Schindler}}, \bibinfo {author} {\bibfnamefont {D.}~\bibnamefont {Nigg}},
  \bibinfo {author} {\bibfnamefont {T.}~\bibnamefont {Monz}}, \bibinfo {author}
  {\bibfnamefont {M.}~\bibnamefont {Chwalla}}, \bibinfo {author} {\bibfnamefont
  {M.}~\bibnamefont {Hennrich}}, \bibinfo {author} {\bibfnamefont {C.~F.}\
  \bibnamefont {Roos}}, \bibinfo {author} {\bibfnamefont {P.}~\bibnamefont
  {Zoller}}, \ and\ \bibinfo {author} {\bibfnamefont {R.}~\bibnamefont
  {Blatt}},\ }\href@noop {} {\bibfield  {journal} {\bibinfo  {journal}
  {Nature}\ }\textbf {\bibinfo {volume} {470}},\ \bibinfo {pages} {486}
  (\bibinfo {year} {2011})}\BibitemShut {NoStop}%
\bibitem [{\citenamefont {Reiter}\ and\ \citenamefont
  {S\o{}rensen}(2012)}]{Reiter12}%
  \BibitemOpen
  \bibfield  {author} {\bibinfo {author} {\bibfnamefont {F.}~\bibnamefont
  {Reiter}}\ and\ \bibinfo {author} {\bibfnamefont {A.~S.}\ \bibnamefont
  {S\o{}rensen}},\ }\href {\doibase 10.1103/PhysRevA.85.032111} {\bibfield
  {journal} {\bibinfo  {journal} {Phys. Rev. A}\ }\textbf {\bibinfo {volume}
  {85}},\ \bibinfo {pages} {032111} (\bibinfo {year} {2012})}\BibitemShut
  {NoStop}%
\bibitem [{\citenamefont {Friis}\ \emph {et~al.}(2017)\citenamefont {Friis},
  \citenamefont {Marty}, \citenamefont {Maier}, \citenamefont {Hempel},
  \citenamefont {Jurcevic}, \citenamefont {Plenio}, \citenamefont {Huber},
  \citenamefont {Roos}, \citenamefont {Blatt}, \citenamefont {Lanyon} \emph
  {et~al.}}]{friis17}%
  \BibitemOpen
  \bibfield  {author} {\bibinfo {author} {\bibfnamefont {N.}~\bibnamefont
  {Friis}}, \bibinfo {author} {\bibfnamefont {O.}~\bibnamefont {Marty}},
  \bibinfo {author} {\bibfnamefont {C.}~\bibnamefont {Maier}}, \bibinfo
  {author} {\bibfnamefont {C.}~\bibnamefont {Hempel}}, \bibinfo {author}
  {\bibfnamefont {P.}~\bibnamefont {Jurcevic}}, \bibinfo {author}
  {\bibfnamefont {M.}~\bibnamefont {Plenio}}, \bibinfo {author} {\bibfnamefont
  {M.}~\bibnamefont {Huber}}, \bibinfo {author} {\bibfnamefont
  {C.}~\bibnamefont {Roos}}, \bibinfo {author} {\bibfnamefont {R.}~\bibnamefont
  {Blatt}}, \bibinfo {author} {\bibfnamefont {B.}~\bibnamefont {Lanyon}},
  \emph {et~al.},\ }\href@noop {} {\bibfield  {journal} {\bibinfo  {journal}
  {arXiv preprint arXiv:1711.11092}\ } (\bibinfo {year} {2017})}\BibitemShut
  {NoStop}%
\bibitem [{\citenamefont {Bertini}\ \emph
  {et~al.}(2016{\natexlab{b}})\citenamefont {Bertini}, \citenamefont {Collura},
  \citenamefont {De~Nardis},\ and\ \citenamefont {Fagotti}}]{bertini16a}%
  \BibitemOpen
  \bibfield  {author} {\bibinfo {author} {\bibfnamefont {B.}~\bibnamefont
  {Bertini}}, \bibinfo {author} {\bibfnamefont {M.}~\bibnamefont {Collura}},
  \bibinfo {author} {\bibfnamefont {J.}~\bibnamefont {De~Nardis}}, \ and\
  \bibinfo {author} {\bibfnamefont {M.}~\bibnamefont {Fagotti}},\ }\href
  {\doibase 10.1103/PhysRevLett.117.207201} {\bibfield  {journal} {\bibinfo
  {journal} {Phys. Rev. Lett.}\ }\textbf {\bibinfo {volume} {117}},\ \bibinfo
  {pages} {207201} (\bibinfo {year} {2016}{\natexlab{b}})}\BibitemShut
  {NoStop}%
\bibitem [{\citenamefont {Castro-Alvaredo}\ \emph {et~al.}(2016)\citenamefont
  {Castro-Alvaredo}, \citenamefont {Doyon},\ and\ \citenamefont
  {Yoshimura}}]{castro-alvaredo16}%
  \BibitemOpen
  \bibfield  {author} {\bibinfo {author} {\bibfnamefont {O.~A.}\ \bibnamefont
  {Castro-Alvaredo}}, \bibinfo {author} {\bibfnamefont {B.}~\bibnamefont
  {Doyon}}, \ and\ \bibinfo {author} {\bibfnamefont {T.}~\bibnamefont
  {Yoshimura}},\ }\href {\doibase 10.1103/PhysRevX.6.041065} {\bibfield
  {journal} {\bibinfo  {journal} {Phys. Rev. X}\ }\textbf {\bibinfo {volume}
  {6}},\ \bibinfo {pages} {041065} (\bibinfo {year} {2016})}\BibitemShut
  {NoStop}%
\bibitem [{\citenamefont {Bulchandani}\ \emph {et~al.}(2017)\citenamefont
  {Bulchandani}, \citenamefont {Vasseur}, \citenamefont {Karrasch},\ and\
  \citenamefont {Moore}}]{bulchandani17}%
  \BibitemOpen
  \bibfield  {author} {\bibinfo {author} {\bibfnamefont {V.~B.}\ \bibnamefont
  {Bulchandani}}, \bibinfo {author} {\bibfnamefont {R.}~\bibnamefont
  {Vasseur}}, \bibinfo {author} {\bibfnamefont {C.}~\bibnamefont {Karrasch}}, \
  and\ \bibinfo {author} {\bibfnamefont {J.~E.}\ \bibnamefont {Moore}},\ }\href
  {\doibase 10.1103/PhysRevLett.119.220604} {\bibfield  {journal} {\bibinfo
  {journal} {Phys. Rev. Lett.}\ }\textbf {\bibinfo {volume} {119}},\ \bibinfo
  {pages} {220604} (\bibinfo {year} {2017})}\BibitemShut {NoStop}%
\end{thebibliography}
%

\end{document}